\documentclass{aastex}
\usepackage{emulateapj5}

\gdef\kms{km s$^{-1}$} 
\gdef\etal{{et al.}}

\gdef\Sec{${}^{\prime\prime}$\llap{.}}

\gdef\etal{{et~al.}}

\gdef\kms{{km~s$^{-1}$}}
\gdef\kpc-1{{kpc$^{-1}$}}
\gdef\Mpc-1{{Mpc$^{-1}$}}
\gdef\s-1{{sec$^{-1}$}}
\gdef\pdeg2{{deg$^{-2}$}}

\gdef\h0{{H$_0$}}
\gdef\q0{{$q_0$}}

\gdef\rms{{\it rms\/}}

\gdef\kms{\hbox{$\rm km\,s^{-1}$}}

\gdef\ltsima{$\scriptscriptstyle \; \buildrel < \over \sim \;$}
\gdef\simlt{\lower.3ex\hbox{\ltsima}}

\gdef\gtsima{$\scriptscriptstyle \; \buildrel > \over \sim \;$}
\gdef\simgt{\lower.3ex\hbox{\gtsima}}

\gdef\about{\raise.3ex\hbox{$\scriptscriptstyle \sim $}}

\gdef\Sec{\hbox{${}^{\prime\prime}$\llap{.}}}

\gdef\sqr#1#2{{\vcenter{\vbox{\hrule height.#2pt
        \hbox{\vrule width.#2pt height#1pt \kern#1pt
        \vrule width.#2pt}
        \hrule height.#2pt}}}}

\shorttitle{The Dark Matter Profile of Abell 2199}

\slugcomment{Accepted for Publication in the Astrophysical Journal}

\begin{document}

\title{DETERMINATION OF THE DARK MATTER PROFILE OF ABELL 2199 FROM 
INTEGRATED STARLIGHT\altaffilmark{1}}

\author{Daniel D. Kelson\altaffilmark{2,3},
Ann I. Zabludoff\altaffilmark{4},
K.A. Williams\altaffilmark{5},
S.C. Trager\altaffilmark{6,7},
J.S. Mulchaey\altaffilmark{6},
and
Michael Bolte\altaffilmark{5}
}

\altaffiltext{1}{Based on observations obtained at the W. M. Keck
Observatory, which is operated jointly by the California Institute of
Technology and the University of California.}

\altaffiltext{2}{Department of Terrestrial Magnetism, Carnegie
Institution of Washington, 5241 Broad Branch Rd., NW, Washington, DC
20015}

\altaffiltext{3}{The Observatories of the Carnegie Institution of
Washington, 813 Santa Barbara Street, Pasadena, CA 91101}

\altaffiltext{4}{Steward Observatory, University of Arizona, Tucson, AZ
85721}

\altaffiltext{5}{UCO/Lick Observatory, Department of Astronomy and
Astrophysics, University of California, Santa Cruz, CA 95064}

\altaffiltext{6}{The Observatories of the Carnegie Institution of
Washington, 813 Santa Barbara Street, Pasadena, CA 91101}

\altaffiltext{7}{Hubble Fellow}

\begin{abstract}

We have obtained deep, long-slit spectroscopy along the major axis of
NGC 6166, the cD galaxy in the cluster Abell 2199, in order to measure
the kinematics of intracluster starlight at large radii. The velocity
dispersion initially decreases from the central value of 300 \kms, to
200 \kms\ within a few kpc, and then steadily rises to 660 \kms at a
radius of 60 kpc (H$_0$=75 \kms\Mpc-1, $\Omega_m$=0.3,
$\Omega_\Lambda$=0.7), nearly reaching the velocity dispersion of the
cluster ($\sigma_{\rm A2199}=775\pm 50$ \kms). These data suggest that
the stars in the halo of the cD trace the potential of the cluster and
that the kinematics of these intracluster stars can be used to constrain
the mass profile of the cluster. In addition, we find evidence for
systematic rotation ($V/\sigma$$\approx$0.3) in the intracluster stars
beyond 20 kpc. Such rotation is not seen in the kinematics of the
cluster members.

The surface brightness and velocity dispersion profiles can be fit using
a single component mass model only by making unphysical assumptions
about the level of anisotropy for both the stars in the cD galaxy and
for the kinematics of the galaxies in the cluster. Two-component mass
models for the cD and its halo are subsequently explored using the
kinematics of known cluster members as an additional constraint on the
total enclosed mass beyond the extent of the stellar kinematics. Under
the assumption of isotropy, the observed major-axis kinematics can be
reproduced only if the halo, parameterized by a generalized-NFW profile,
has a soft core, {\it i.e.}, $\alpha < 1$ (a generalized-NFW halo with
$\alpha=1$ is excluded due to low implied stellar mass-to-light ratios).
This result is inconsistent with the predictions of current $N$-body
simulations for dark matter halos.

To test the consistency of our halo profiles with those derived from
strong lensing measurements in intermediate redshift clusters, we
calculate the critical radii for tangential arcs, assuming that our
best-fit mass models for the Abell 2199 were placed at cosmological
redshifts between $0.2 \le z\le 0.5$. The calculated critical radii for
our best-fit two-component isotropic models range from $5''$ to $40''$,
depending on the assumed source redshift, consistent with the radii for
gravitational arcs observed in intermediate redshift clusters.

We also present the results of Monte Carlo simulations testing the
general reliability of velocity dispersion measurements in the regime of
low $S/N$ and large intrinsic Doppler broadening.

\end{abstract}

\keywords{ galaxies: elliptical and lenticular, cD, galaxies:
kinematics and dynamics, galaxies: stellar content, galaxies:
clusters: individual (Abell 2199), galaxies: individual (NGC 6166),
(cosmology:) dark matter}

%%%%%%%%%%%%%%%%%%%%%%%%%%%%%%%%%%%%%%%%%%%%%%%%%%%%%%%%%%%%%%%%%%%%%%%%

\begin{figure*}[ht]
\epsscale{0.8}
\plotone{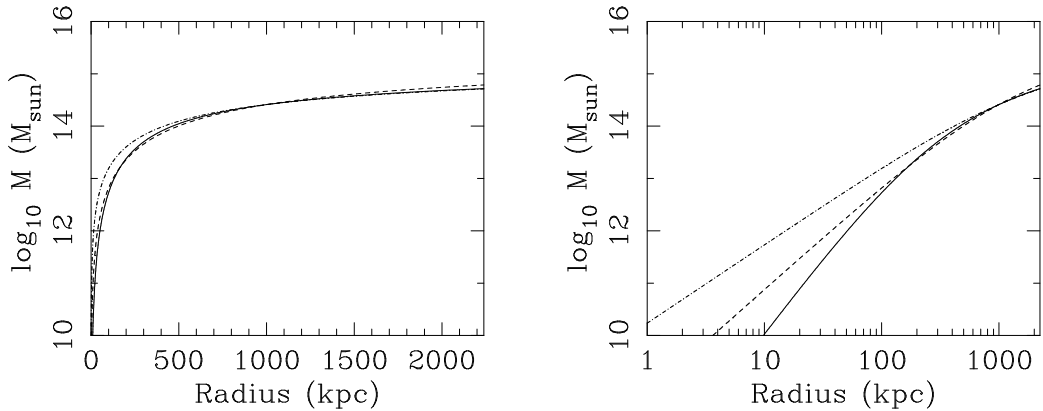}
\figcaption[fig1.eps]{Mass profiles for generalized NFW dark matter halos with
$\alpha\in {0,1,1.5}$ (solid,dashed,dot-dashed, respectively). These mass profiles
have been renormalized to have identical mass at 1 Mpc. The left and
right panels are identical, except for the linear and logarithmic
stretch in radius. Note the range of projected radii, $R \simlt 100$
kpc, over which one would observe significant differences.
\label{samples}}
\end{figure*}

\section {Introduction}
\label{intro}

The question of how mass is distributed in clusters is long-standing
\citep{smith,smith2}. Cosmological $N$-body simulations attempt to
address this question and, despite uncertainties in the input physics,
robustly predict a universal mass density profile for dark matter halos
\citep{nfw,moore,ghigna}. The spherically symmetric
density profile is parameterized by \begin{equation}
\rho \propto {1 \over x^\alpha (x+1)^{3-\alpha}}
\label{eq:nfw}
\end{equation}
where $x=r/r_s$, and $r_s$ is the scale radius at which the slope of the
density profile transitions to the steep $r^{-3}$ fall-off. Early
studies of the universality of the profile suggested $\alpha=1$
\citep{nfw}. While there remains some controversy, more recent
simulations indicate $\alpha = 1.5$ \citep{ghigna}. The predictions are
based on collisionless $N$-body simulations, but some attempts to
incorporate the effects of gas dynamics have led to steeper inner
profiles \citep[$\alpha \simlt 2$;][]{tissera,alvarez}. Alternative forms
of dark matter have also been explored
\citep[e.g.,][]{spergel,firmani1,colin}. Halos of self-interacting dark matter
have soft, uniform density cores \citep[$\alpha<1$;][]{yoshida2,romeel},
but their low ellipticities are considered problematic
\citep[e.g.,][]{jordi}.

%to reconcile observations (e.g.,
%\citep{swaters} discrepant with the simulations (e.g.,
%\citep{klypin,jing}.

Some CDM simulations indicate that $\alpha$ itself may be a function of
halo mass, with cluster-sized halos tending to $\alpha=1$ and
galaxy-sized halos tending to $\alpha=1.5$ \citep[][though see
\cite{klypin}]{jing}. Such results suggest that the density
profiles of dark matter halos are quite sensitive to their histories of
formation, accretion, and merging, as well as to the initial conditions
of collapse \citep{nusser}, and these notions are tested explicitly by
Weinberg (2001), who finds that noisy evolutionary processes can drive
halo profiles to a uniform shape with soft cores. This result appears
similar to the profile derived by Iliev \& Shapiro (2001), wherein the
collapse of spherical top-hat density perturbations was shown to produce
truncated isothermal spheres, i.e., $\alpha=0$. The solutions of Iliev
\& Shapiro (2001) and Weinberg (2001) may naturally produce soft cores
because of the boundary condition of finite phase-space densities at
$r=0$. This restriction on the solutions to the Lame-Emden or
Fokker-Planck equations \citep{iliev,weinberg} runs counter to the basic
assumption in fitting Eq. \ref{eq:nfw} to $N$-body profiles, wherein the
central phase-space densities are allowed to diverge \citep{taylor}.

While such analytic work is providing insight into the physics of dark
matter halo evolution, there are few observational constraints on
cluster scales to test the $1\le \alpha \le 1.5$ models routinely
assembled by the simulations. For the massive dark matter halos of rich
clusters, this range in $\alpha$ yields models that are quite similar
outside of $\simlt 100$ kpc, where most cluster-related observations
have weight. Figure \ref{samples} shows the mass profiles of example
generalized-NFW mass profiles for $\alpha\in \{0,1,1.5\}$. The two
panels show the mass profiles using linear and logarithmic stretches in
cluster radius and illustrate the need for observations within the inner
$\simlt 100$ kpc in order to measure the inner slope of the dark matter
halo profile \citep[see also][]{klypin}.

\begin{figure*}[ht]
\epsscale{1.8}
%\plotone{fig2.eps}
\vspace{4in}
\figcaption[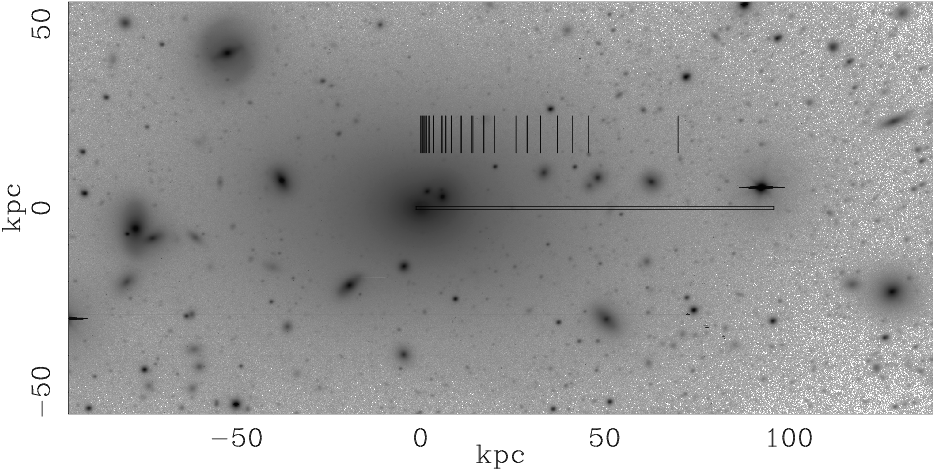]{A $\log$-stretch of a $B$-band image of NGC 6166 with the
position of the LRIS long-slit shown for the primary integrations
(totaling 7200 s). The radial bins, from which the galaxy spectra were
extracted, are shown. Note the increasingly large apertures from which
the two-dimensional spectra were summed to achieve adequate
signal-to-noise ratio. A second long-slit position, exposed for 1400 s,
was also obtained with the slit centered on the galaxy nucleus.
\label{finder}}
\end{figure*}

Observational tests at these small radii are scarce. The determination
of the mass profile from kinematics of cluster galaxies within this
region has been difficult because there are generally few members. van
der Marel \etal\ (2000) attempted to solve this problem by combining the
redshifts from the CNOC1 cluster survey \citep{cnoc1} into a single
fiducial cluster. They argued that the best-fit mass profiles were
described by $\alpha\approx 1$, but their analysis did not take into
account the contribution of the stellar mass of any brightest cluster
galaxies (BCG) to the total mass profile of the fiducial cluster. An
accurate determination of mass profiles of the inner 100 kpc for
clusters from X-ray observations has been hampered to date by poor
spatial resolution, and by the presence of cooling flows and shocked
regions. Even as X-ray observations improve, uncertainties about the
mass contributions from BCGs will persist \citep[see][]{arabadjis}.
Still another method, that of using the kinematics of planetary nebulae
and/or globular clusters \citep[e.g.,][]{cohen,cote}, requires
painstaking identification and spectroscopy of 20-200 discrete objects,
and Hernquist \& Bolte (1993) show that the resulting inferred mass
profiles are subject to large ($>$ factors of 2) uncertainties with such
small samples. Strong gravitational lensing measurements can allow one
to infer the total surface mass density within the Einstein radius but
these constraints are susceptible to degeneracies \citep{gorenstein} and
to uncertainties in the details of the potential \citep{bartelmann}. The
mass profile can be better-constrained when there are several arcs, but
the brightest cluster galaxies may themselves dominate the shapes of the
total surface mass density profile, as may be the case in MS2137-2353
\citep{hammer}. Weak lensing measurements are even less effective at
constraining the inner mass profile, given the typical number densities
of background galaxies as well as the comparatively large surface areas
covered by the galaxies in cluster cores \citep{hoekstra}.

A more promising method for constraining the mass profiles in the cores
of rich clusters is to use the kinematics of stars in the extended halos
of central cD galaxies (``intracluster starlight''). Past efforts to
trace the stellar velocity dispersion profile of IC 1101, in the cluster
A2029, to tens of kpc showed that $\sigma(r)$ rises from 400 \kms\ to
600 \kms\ towards larger radii \citep{dressler,carter,tonry2}. Of the
many BCGs investigated over radii of $10''-20''$, only IC1101
\citep{fisher,kronawitter} and NGC 6166 \citep{carter1999} have shown
such a rise. Such results suggest that the stars in a cD's halo
respond to the potential of the cluster.

In this paper we report on a pilot study in which we targeted NGC 6166,
the giant elliptical in the center of the rich cluster Abell 2199 ($z =
0.03$), with the Keck II 10m telescope and the Low Resolution Imaging
Spectrograph \citep[][; LRIS]{okelris}. Several factors make this galaxy
an ideal choice: (1) it is at rest with respect to the center of the
cluster potential \citep{ann90}; (2) the cluster itself appears relaxed
in X-rays; (3) the presence of a cooling flow in the cluster core
suggests little or no recent merger activity in the cluster center; (4)
the useful features in its spectrum (e.g., the G-band) do not overlap
any strong night-sky emission lines or absorption lines; (5) the
redshift allows the $170''$ LRIS slit to subtend $100$ kpc ($H_0= 75$
\kms\ Mpc$^{-1}$, $\Omega_M=0.3$, and $\Omega_\Lambda=0.7$; used
throughout this paper); and (6) its declination of $39^{\circ}$ makes it
observable for most of the night from Mauna Kea. The telescope's large
aperture makes it possible to obtain spectra with sufficiently high
$S/N$ ratios in the low surface brightness halo of the galaxy while
using a narrow slit-width (both to achieve moderate resolution and to
exclude interloping galaxies).

We structure the paper as follows. In \S \ref{data} we discuss the
observations and data reduction. The stellar mass density profile is
constrained in \S \ref{stellarmass} through the use of the $B$- and $R$-
band surface brightness profiles, and the major-axis kinematics of NGC
6166 are derived in \S \ref{kinematic}. In \S \ref{montecarlo} we
discuss the reliability of velocity dispersion measurements in the
regime of low $S/N$ and large intrinsic Doppler broadening using Monte
Carlo simulations. Section \ref{models} discusses the fit of simple mass
models to the kinematics of NGC 6166 and Abell 2199, as constrained by
the kinematics of cluster members. The implications of our best-fit mass
models are discussed in \S \ref{discuss} and our conclusions are
summarized in \S \ref{summary}.

%and discuss the implications of our models in the contexts of
%other estimates for the mass profile of Abell 2199, and analyses of
%distant clusters using strong gravitational lensing.

%%%%%%%%%%%%%%%%%%%%%%%%%%%%%%%%%%%%%%%%%%%%%%%%%%%%%%%%%%%%%%%%%%%%%%%%

\begin{figure*}[ht]
\plotone{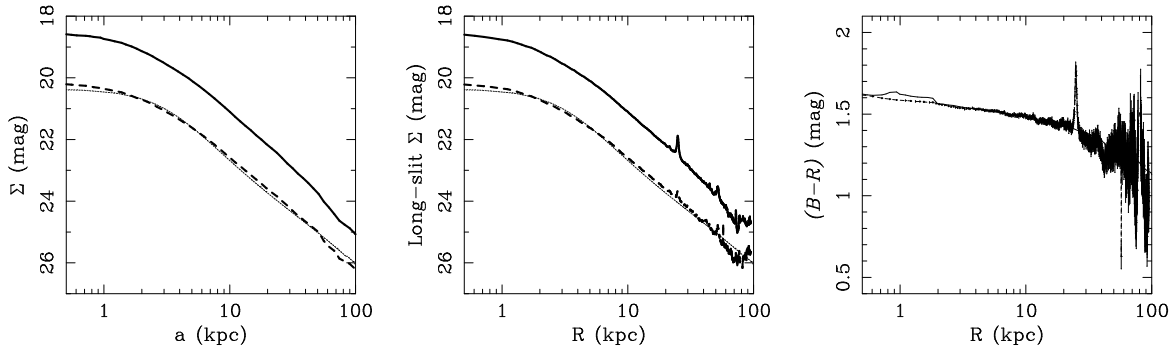}
\figcaption[fig3.eps]{(left) The mean $R$- (solid) and $B$-band (dashed)
surface brightness profile of NGC 6166, derived from isophote fitting,
as a function of the major-axis of the isophote. (middle) The $R$- and
$B$-band surface brightness profile of NGC 6166, derived along the
position of the LRIS long-slit. The dotted line in both panels indicates
the King model that best fits the $B$-band profile as measured along
the position of the long-slit (see \S \ref{models}. The fit of the
King profile was restricted to radii $2\Sec 5<r< 125''$ (1.2 kpc $<r<$
70 kpc) and satisfactorily matches the profiles determined from the
isophotes and from the long-slit position. (right) The $B-R$ color
gradients derived from the isophote fitting (solid line) and the
long-slit position (dashed line). Note that the cD becomes
bluer with projected radius. The color gradient is indicative of a
radial variation in the metallicities and/or ages of the stellar
populations and implies that the stellar $M/L$ ratio is also varying
with projected radius.
\label{sb_prof}}
\end{figure*}

\section{Observations and Data Reduction}
\label{data}

The primary goal of this work is the accurate measurement of the second
moment ($\sigma$) of the line-of-sight velocity distribution (LOSVD) as
a function of position along the major-axis of NGC 6166. By combining
the $\sigma$ profile with broad-band imaging, we can obtain constraints
on the total mass-to-light ($M/L$) profile of the galaxy and the inner
portions of the surrounding cluster. In this section we describe the
$B$- and $R$-band imaging of NGC 6166 and its halo, the long-slit
spectroscopy, and the processing required to derive the chief products:
the surface brightness profile, $\mu(R)$, and the projected
line-of-sight velocity dispersion profile, $\sigma(R)$. Using the
cosmological parameters given in \S \ref{intro}, the scale on the sky is
$0.56$ kpc/arcsec.

%%%%%%%%%%%%%%%%%%%%%%%%%%%%%%%%%%%%%%%%%%%%%%%%%%%%%%%%%%%%%%%%%%%%%%%%
\subsection{The Photometry}

Three $B$-band images of NGC 6166, totaling 1870 seconds, and six
$R$-band images, totaling 1280 seconds, were obtained with LRIS during
the nights of 30-31 May 1998. The images span $6\times 8$ square arcmin,
corresponding to a distance along the major axis of 170 kpc. The seeing
during the run was approximately 0.8 arcsec (FWHM). A subsection of the
$B$-band image is shown in Figure \ref{finder}.

\subsubsection{The Surface Brightness Profile}

The sky background was determined for each image using the corners of
the image furthest from the galaxy, and the sky brightness was then
subtracted from the entire image (in both bands). While errors in the
sky will severely affect the derived surface brightness profile of the
galaxy at radii approaching the extent of the imaging field-of-view, the
accuracy in our estimate of the sky is suitable for deriving the surface
brightness profile over the extent of the LRIS long-slit (\about $3'$).
The $B$- and $R$-band sky brightnesses were approximately $\mu_{B,sky}
=21.9$ mag/arcsec$^2$ and $\mu_{R,sky} =20.6$ mag/arcsec$^2$,
respectively.

The fitting of elliptical isophotes was performed using the {\sc
ellipse\/} package of STSDAS in IRAF. The detection of 15 interloping
galaxies necessitated modeling (using the ellipse package) and
subtraction of their light from the images in both bands. Isophotes were
fit to these galaxies, as well as to the cD, and iteratively subtracted
from the image. The envelope of one interloper intersects the major-axis
between 2 kpc and 14 kpc at a mean level of about 15\% of the cD surface
brightness. The residuals from the subtraction of this particular
interloper amounted to \about 5\% of the cD surface brightness at those
radii along the major-axis. Such residuals were efficiently rejected by
the sampling criteria in the final ellipse-fitting of the cD.

\subsubsection{The Calibration}

We calibrated the $B$-band photometry using secondary standard stars in
the Galactic globular cluster M92, with an error in the photometric
zero-point of $\pm 0.016$ mag.  Calibrating the $R$-band photometry was
problematic because the LRIS filter wheel had not properly inserted the
(Johnson) $R$ filter into the optical path for the M92 calibration
image. As a result we used the Cousins $R$-band surface brightness
profile from \cite{postman} to calibrate our data. The \cite{postman}
profile was first transformed from Cousins $R$ to Johnson $R$ using
$R_J-R_C = -0.11$ (a good approximation for the spectral energy
distributions of old E/S0s; \citep{fuk95}. We then determined the
photometric offset between the LRIS profile and the transformed
\cite{postman} profile over the radial range of 5 to 40 arcsec. The
$rms$ of this offset was $0.017$ mag.

The $B$-band extinction to NGC 6166, as estimated by Schlegel \etal\
(1998), is $A_B=0.05$ mag \citep[$A_B=0.00$ mag from ][]{bh82}. Because
this prediction is quite low and irrelevant for the determination of the
mass profile we have made no correction for the foreground extinction
(it only affects the deduced stellar $M/L$ ratios at a level of
$\simlt 5\%$). The final $B$- and $R$-band surface brightness profiles
are shown in the left panel of Figure \ref{sb_prof}.

\subsubsection{The Surface Brightness Profile along the LRIS Long-Slit}

While isophote fitting provides the maximum signal-to-noise ratio in
the surface brightness profile at a given radius, our kinematic profile
will be derived along a fixed position angle (the major-axis). Thus it
is more appropriate to use the surface brightness profile along this
position angle in modeling the mass profile and constraining the mass
model of the cD and its halo.

We calculated the $B$- and $R$-band slit profiles from eight columns of
the CCD image centered on the galaxy nucleus. The long-slit intensities
were derived by averaging the eight columns of the slit position in
order to produce a single intensity per pixel for each row along the
slit. These $B$- and $R$-band long-slit surface brightness profiles are
shown in the middle panel of Figure \ref{sb_prof}. The profiles in the
right- and left-hand panels are virtually identical because there was
very little twisting of the isophotes with radius. Thus our mass
modeling will be insensitive to the choice of the long-slit or isophote
profile.  For the remainder of the paper, we use the profile derived at
the position of the LRIS long-slit.

\begin{figure*}[t]
\epsscale{1.0}
\plotone{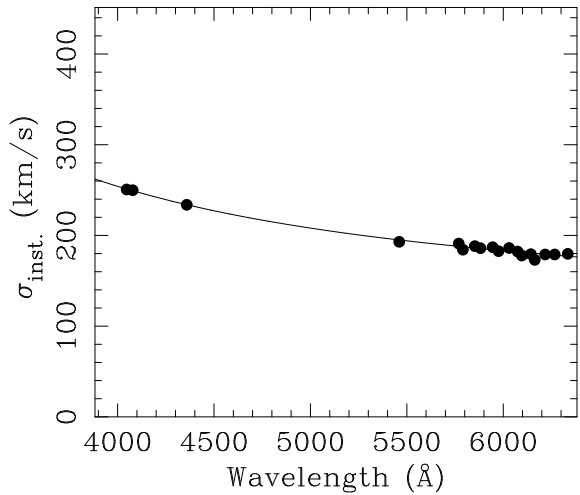}
\epsscale{1.0}
\figcaption[fig4.eps]{The instrumental resolution for the LRIS configuration
used for the work presented in this paper. The resolution was determined
from the widths of the emission lines from the mercury, krypton, and
neon lamps.
\label{resol}}
\end{figure*}

\subsubsection{The Color Gradient in NGC 6166}
\label{colorgrad}

The right-hand panel of Figure \ref{sb_prof} shows the $B-R$ color
gradients derived from both the isophote fitting (solid line) and for
the long-slit position (dashed line). The color transitions smoothly
from $B-R=1.6$ mag in the inner regions of the cD to $B-R=1.4$ mag at
$r=20$ kpc to $B-R=1.2$ mag at $r=50$ kpc. These are colors which
are typical for E/S0, Sa, and Sbc galaxies, respectively \citep{fuk95}.
We find that the gradient is well fit by the second-order polynomial:
\begin{equation}
(B-R)= 1.58 -0.10 [\log r {\rm (kpc)}]^2
\end{equation}
with an \rms\ scatter of 0.018 mag.

\smallskip

In \S \ref{stellarmass} we derive structural parameters from the
surface photometry and discuss the use of the color gradient to estimate
possible variations in stellar $M/L$ with projected radius.

%%%%%%%%%%%%%%%%%%%%%%%%%%%%%%%%%%%%%%%%%%%%%%%%%%%%%%%%%%%%%%%%%%%%%%%%

\subsection{The Spectroscopy}

The primary spectroscopic data consist of four 40 minute exposures of
the galaxy, flanked by five 40 minute offset-sky exposures (taken before
and after the on-galaxy exposures). We used the $1\Sec 5$ wide long-slit
with the 600 grooves mm$^{-1}$ grating, blazed at 5000\AA. The spectral
coverage extends from 3800\AA\ to 6400\AA. The resolution of \about
7\AA\ FWHM is equivalent to $\sigma_{\rm inst.}\approx 200$ km/s, as
shown in Figure \ref{resol}, more than adequate to resolve the velocity
dispersions in the cD and its halo. While the spectral coverage is
suitable for deriving absorption line velocity dispersions from several
features, {\it e.g.,} the Mg b, G-band, or Ca H \& K, we concentrated
our efforts around the G-band as a compromise between the desires to
minimize contamination due to the sky background and to achieve adequate
signal-to-noise ratios.

\begin{figure*}[t]
\epsscale{1.3}
\plotone{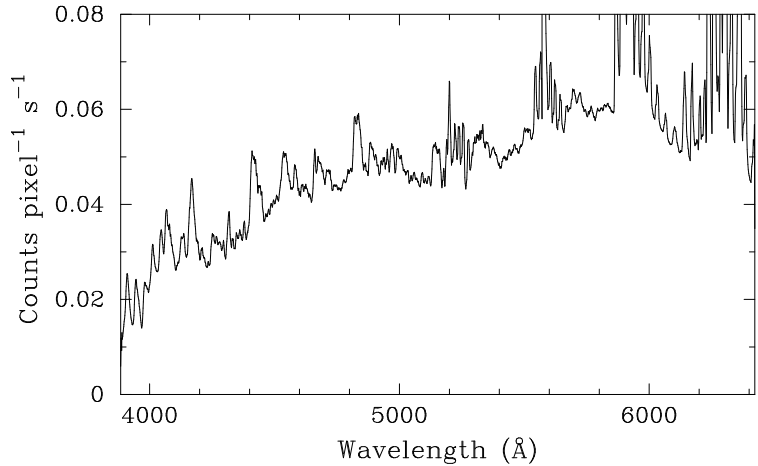}
\epsscale{1.0}
\figcaption[fig5.eps]{A spectrum of the Mauna Kea night sky in the blue. Note
the tremendous amount of structure blueward of O I 5577\AA, including
the (strongly variable) nitrogen line at 5199\AA, which is superposed on
OH Q \& R lines at \about 5200\AA. Because most of the features are broad
with very sharp edges, poor sky subtraction of low-surface brightness
sources can artificially alter the broad absorption features intrinsic to
a galaxy spectrum.
\label{skyspec}}
\end{figure*}

To maximize the extent of the halo observed and still preserve radial
profile information for NGC 6166, we placed one end of the long-slit at
the galaxy center and the rest of the $170''$ long-slit north-east along
the major axis at a position angle of 33 degrees (see Figure
\ref{finder}. Our imaging of NGC 6166 on the same night confirms this
angle as that of the major axis to within a degree. A second position,
at the same position angle but centered on the galaxy nucleus, was also
obtained in 1400 sec of integration time.

Because NGC 6166 fills the entire length of a long-slit placed long the
major-axis, we obtained alternating sky pointings offset from the
galaxy, similar to the approach used by Dressler (1979), to facilitate
accurate removal of the sky background. In principle, obtaining the sky
spectrum at the same position on the detector as the galaxy spectra
should minimize the effects of imperfect flat fielding, errors in
rectification, and uncertainties in interpolating the sky background
from the slit ends. While Sembach \& Tonry (1996) used charge shuffling
and chopping of the telescope on 5 min timescales in order to ensure
that the offset-sky spectra were obtained at precisely the same physical
pixel locations as the galaxy spectra, LRIS cannot be used exactly in
this mode. Therefore we chose to nod between galaxy and sky on 40 min
intervals, having to read out the detector at each position/exposure.
Our exposure times were chosen as a compromise between (1) requiring the
data in the low-surface brightness halo of the cD to be limited by the
noise in the sky background and not by the read noise in the
electronics, and (2) avoiding the time-variability of the background.
Despite the variability of the sky spectrum, and the flexure
characteristics of the instrument, we were able to obtain quality halo
spectra where the surface brightness of the galaxy is approximately 6\%
of sky (see below).

\subsubsection{Rectification and Wavelength Calibration}

We used the software package E{\sc xpector\/} \citep{expector} to
rectify and wavelength-calibrate the LRIS spectroscopic data. A
combination of Hg, Kr, Ne, and Ar lamps allowed us to rectify the
spectra and to derive the high order terms for the wavelength
calibration. The $rms$ scatter about the dispersion solution of the lamp
spectra was 0.04\AA\ (3 \kms\ at the G band). The zero-points of the
dispersion solutions of the galaxy spectra were refined using several
high $S/N$ sky emission lines (e.g., 5224\AA, 5238\AA, 5577\AA,
6300\AA), resulting in a final $rms$ scatter of 0.6\AA\ ($\pm 40$ \kms\
at the G-band). This large residual scatter about the sky lines is
typical for LRIS data largely because the lines in lamp spectra are
asymmetric compared with the sky emission lines. The asymmetries in the
line profiles systematically vary with position across the CCD causing
small systematic errors in the measurements of the line centers, and
thus introduce a bias in the wavelength scale \citep{expector}. These
large residuals appear as a moderate-order deviation from the lamp line
dispersion solution and may be reduced with further work if one intends
to utilize a very long spectral baseline \citep[see][]{expector}. If the
direct fitting of broadened templates to the galaxy spectrum is
performed over a relatively short wavelength range, then a moderate
$rms$ in the dispersion solution manifests itself primarily as a
zero-point error, and secondarily as a much smaller effective $rms$
scatter in a local dispersion solution. For our data and fitting range,
the effective scatter is smaller by a factor of four (to $\pm 10$ \kms),
and thus has a negligibly small impact on the measured velocity
dispersions.

\subsubsection{Background Subtraction}

By obtaining full 2D sky spectra at offset positions from the cD, before
and after each galaxy exposure, one should, in principle, be able to
reconstruct the sky background at the time of the galaxy exposures. We
modeled the temporal variability in the sky by fitting low-order
polynomials to the sky values at each pixel, as functions of time and/or
airmass. This method should have produced the expected 2D spectra of the
sky background at the time and/or airmass of the galaxy observations,
but such simple algorithms and interpolations did not allow for
sufficiently accurate subtraction of the variable sky spectrum (even in
the blue; see Figure \ref{skyspec}.

We improved the background subtraction by fitting each CCD row in the
on-galaxy 2D exposures using Equation 8 of \cite{expector}, originally
suggested for measuring velocity dispersions from data from which the
background has not been subtracted. By doing so, each CCD row in the
data was fit as a sum of (1) the high signal-to-noise sky spectrum (from
the offset sky-exposures in the same CCD row), (2) a broadened velocity
dispersion template, and (3) the low-order continuum functions. Thus,
the data were represented by a sum of the galaxy's contribution with the
spectrum of the night sky, scaled by and offset by low-amplitude and
low-order polynomials. This polynomial, equivalent to $<1\%$ errors in
the sky determination, effectively removes the time-variability of the
sky background and any instabilities in the instrument. Additional
broadening of the sky spectrum did not improve $\chi^2$ so we infer that
focus variabilities were negligible. Construction of the background in
this way provides a high signal-to-noise map of the contribution of the
sky spectrum to the data within each row. At large galactic radii, where
the galaxy contributes very little flux, this construction of the
row-to-row map of the background produces a noisy row-to-row map of
galaxy flux. Therefore, the 2D map of the sky contribution was fit by a
low order polynomial in each wavelength bin and this spatially smoothed
map of the sky background was used to perform the background
subtraction. Based on the analysis in \S \ref{montecarlo}, in which we
discuss the robustness of the $\sigma$ measurements, our results are not
sensitive to any remaining uncertainties in the background subtraction.

\begin{figure*}[ht]
\epsscale{1.3}
\plotone{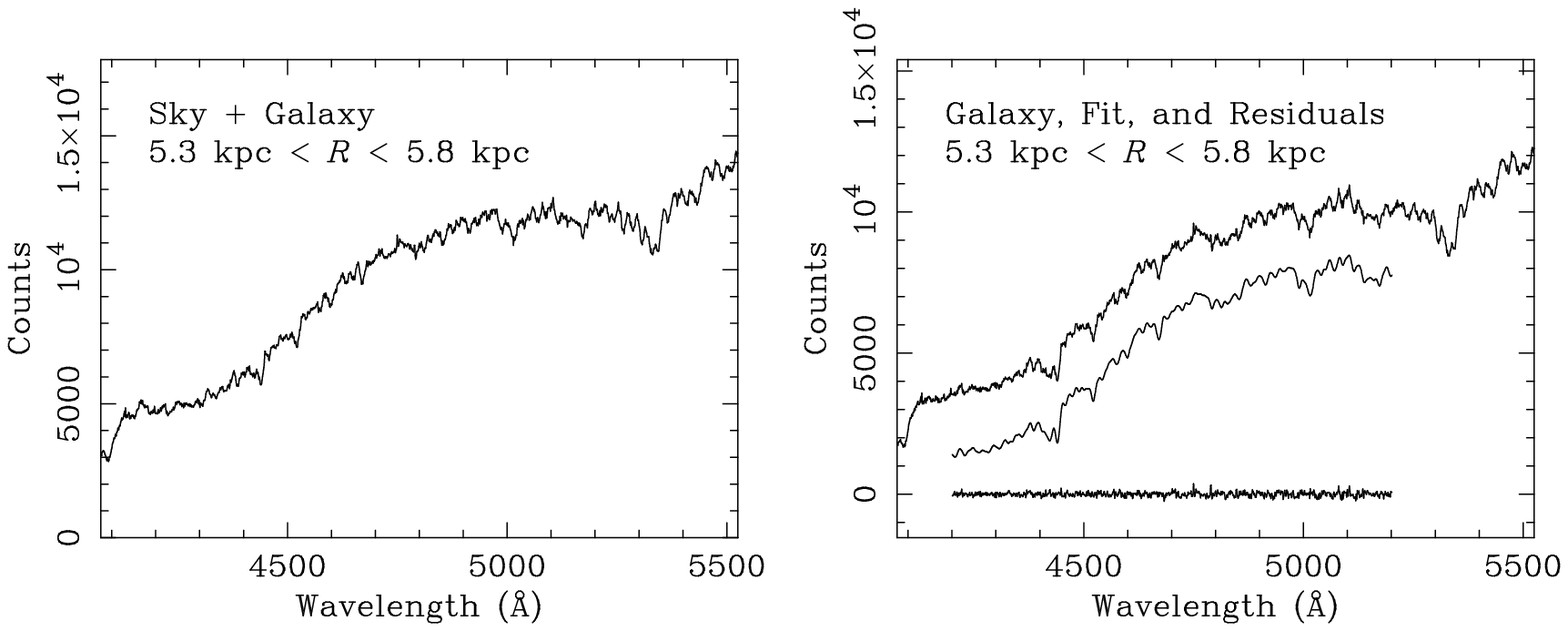}
\plotone{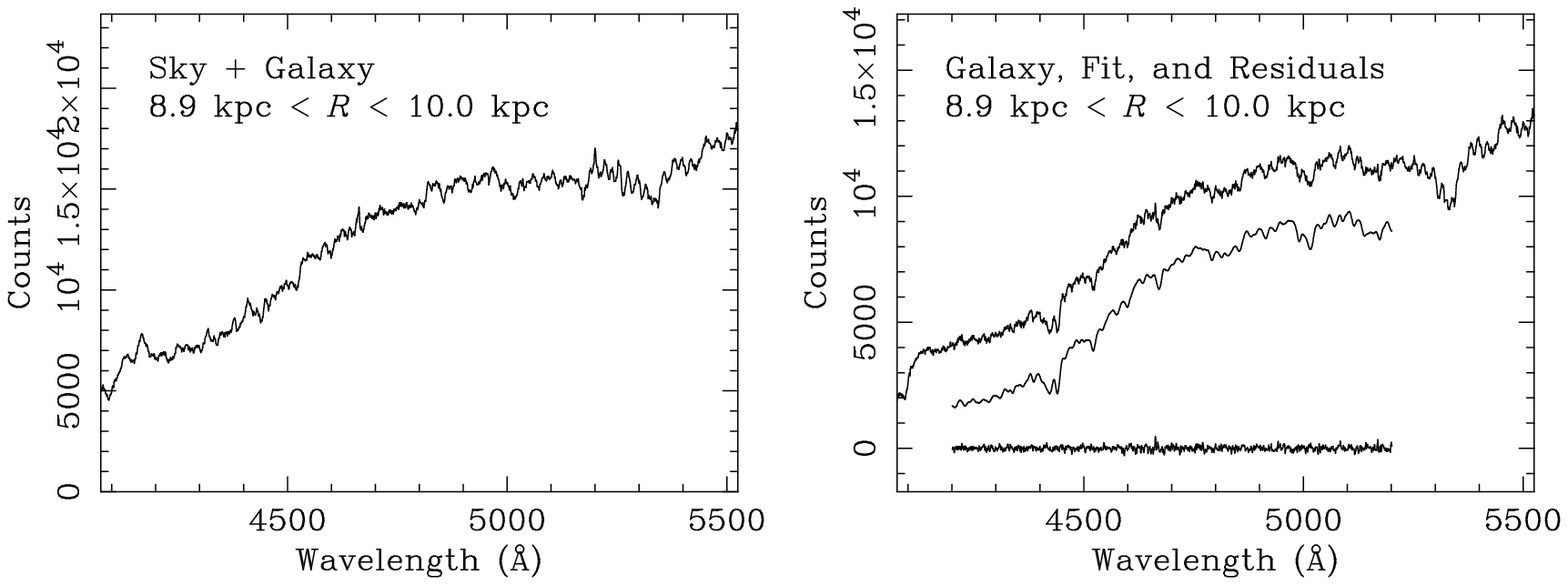}
\plotone{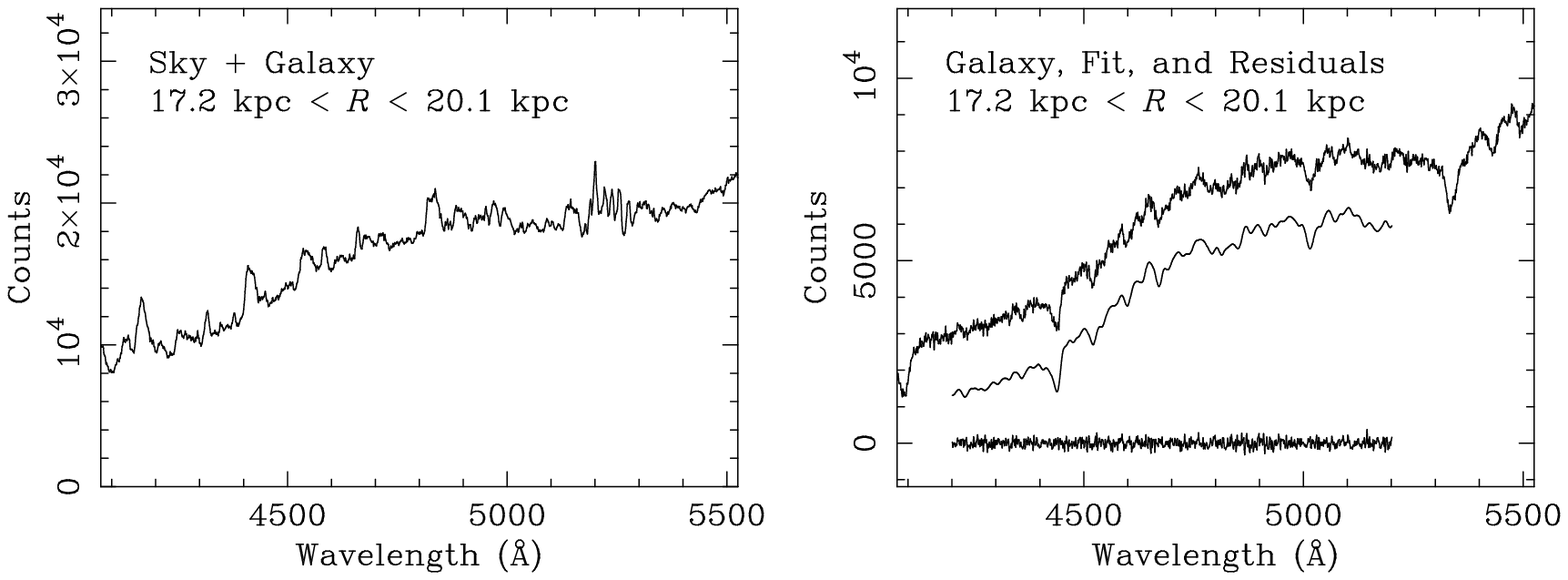}
\plotone{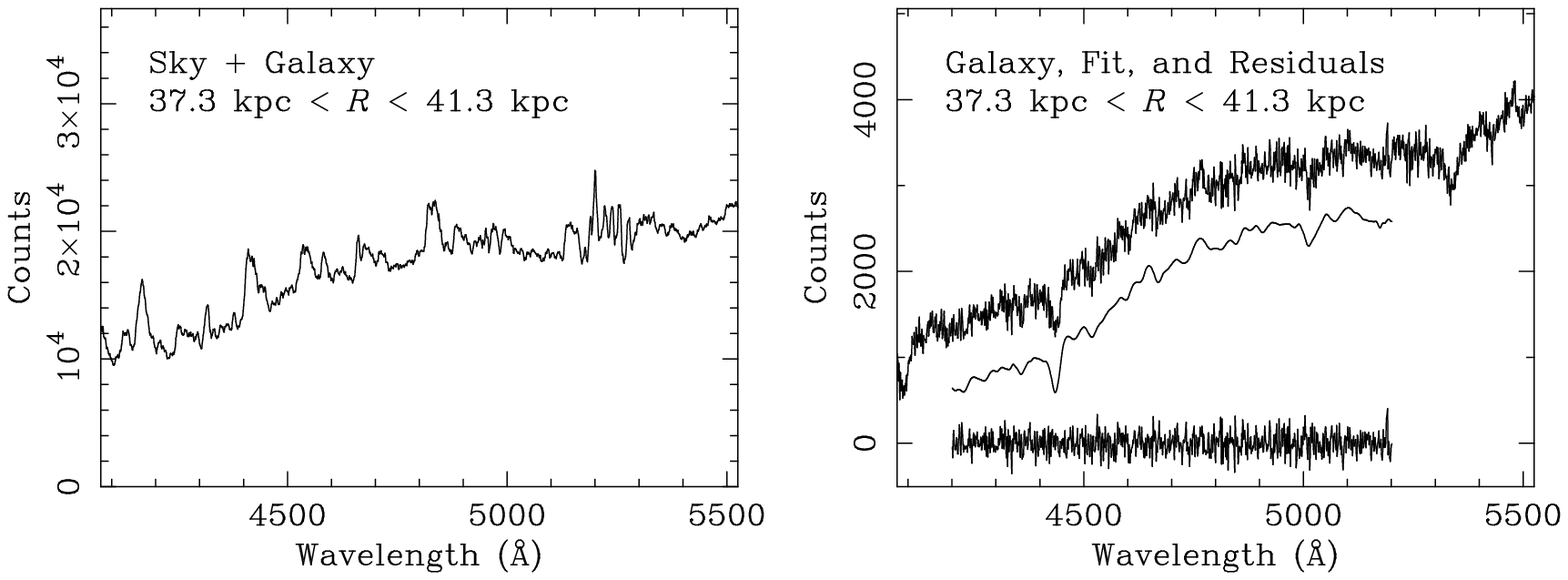}
\epsscale{1.0}
\figcaption[fig6a.eps,fig6b.eps,fig6c.eps,fig7c.eps]{The left-hand
panels show the total sky $+$ galaxy spectra in four representative bins
in projected radius along the major axis of NGC 6166. The right-hand
panels show the galaxy spectra extracted from the same radial bins. The
velocity dispersion fits, shown below the galaxy spectra, illustrate the
quality of the fitting method. The residuals from each fit are also
shown and these indicate that no significant systematic effects remain
unaccounted.
\label{spectra}}
\end{figure*}

\subsubsection{Extraction}
\label{extraction}

Spectra were summed along the spatial direction of the CCD in order to
produce 1D spectra with adequate signal-to-noise ratios ($\ge 20$ per
\AA) for determination of velocity dispersions. Figure \ref{finder}
illustrates the positions of the end-points of the spatial bins using
short solid lines perpendicular to the length of the slit. In Figure
\ref{spectra} we show the total sky$+$galaxy spectra from the deep
spectroscopic integrations of the cD halo in four representative bins
along the major-axis. The corresponding sky-subtracted galaxy spectra
are shown in the right-hand panels of Figure \ref{spectra}. The data
have not been flux-calibrated.

%The $S/N$
%in the sky-subtracted spectra are suitable for measuring velocity
%dispersions to \about 6\% of sky.

\begin{figure*}[b]
\epsscale{1.9}
\plotone{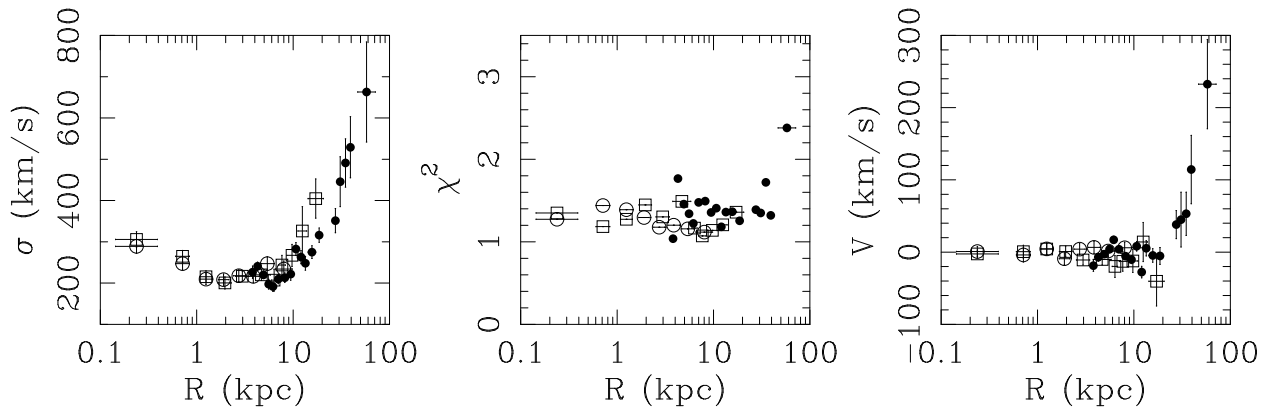}
\epsscale{1.0}
\figcaption[fig7.eps]{(left) The observed projected velocity dispersion
profile of NGC 6166 derived from the deep, off-set long-slit pointing
(filled circles), and from the shallow, central long-slit exposure (open
circles and squares). The squares indicate the profile from the opposite
side of the galaxy from the filled circles, with the projected radii
flipped for comparison. The open and closed circles between 4 kpc
$\simlt R \simlt $ 12 kpc were obtained by simultaneously fitting both
the primary (NGC 6166) and secondary (interloper at +1300 km/s)
component. Note that the velocity dispersion profile has a large
positive gradient in $\sigma$ on both sides of the galaxy after a
decline from the central value of $\sigma_0=300$ km/s. (center) The
reduced $\chi^2$ as a function of position. Our methods provide
excellent fits to the NGC 6166 spectra except in the outermost bin,
where the galaxy surface brightness is 6\% of the sky in the $B$-band.
Between 4 kpc $\simlt r \simlt 12$ kpc the low reduced $\chi^2$ values
indicate that the interloping galaxy at $\delta v=+1300$ km/s was
successfully removed in the fitting procedure and had negligible impact
on the velocity dispersions measured for NGC 6166. (right) The rotation
curve from the same data. Note the apparent increase in rotation in the
outskirts of the cD. \label{raw_prof}}
\end{figure*}

%%%%%%%%%%%%%%%%%%%%%%%%%%%%%%%%%%%%%%%%%%%%%%%%%%%%%%%%%%%%%%%%%%%%%%%%
\section{Constraints on the Stellar Luminosity Density}
\label{stellarmass}

In constructing mass models for NGC 6166, the density profile of the
stellar mass component is constrained by the observed surface brightness
profile. The surface brightness profile only constrains the shape of the
stellar luminosity density profile; the stellar $M/L$ ratio ($M/L_*$)
can only be constrained by the kinematics. Traditionally, photometry
from a single bandpass is used, with the assumption of a spatially
invariant $M/L_*$. However, galaxies have gradients in their stellar
populations
\citep[e.g.,][]{peletier90,franx89,davies,jesus,fishgrad,harald,trager}
and by inference must have gradients in $M/L_*$. Fortunately, we have
surface photometry in both $B$ and $R$, and in this section we combine
the $B-R$ color gradient from \S \ref{colorgrad} with simple stellar
population synthesis models to estimate the gradients in the $B$-band
$M/L_*$ ratio ($M/L_B$) and $R$-band $M/L_*$ ratio ($M/L_R$).

While the nature of the color gradient (as a spatial variation in
stellar ages and/or in stellar chemical abundances) remains uncertain,
we can make some simple assumptions to test to what extent the
gradient in the stellar mass-to-light ratio will affect our
conclusions about the profile of gravitational mass in the system.
We explore {\it three\/} extreme cases: (1) no gradient in the
properties of the stellar population; (2) a gradient in [Fe/H]; and (3)
a gradient in stellar ages. Using the Vazdekis \etal\ (1996) models, we
have derived the correlation of $(B-R)$ color with $M/L_B$ and $M/L_R$
at fixed age or at fixed metallicity:
\begin{eqnarray}
\Delta \log (M/L_B)|_{\hbox{fixed age}} &= 0.96 \Delta (B-R)\cr \Delta
\log (M/L_R)|_{\hbox{fixed age}} &= 0.56 \Delta (B-R)
\end{eqnarray}
and
\begin{eqnarray}
\Delta \log (M/L_B)|_{\hbox{fixed [Fe/H]}} &= 2.00 \Delta (B-R)\cr
\Delta \log (M/L_R)|_{\hbox{fixed [Fe/H]}} &= 1.60 \Delta (B-R).
\end{eqnarray}

Combining the above relations with the color gradient in NGC 6166, we
find that the adoption of a constant $M/L_B$ would lead one to
overestimate the stellar mass densities at large radii by factors of
2--6. By incorporating the gradient in stellar populations, we hope
to derive (presumably) more accurate representations of the stellar mass
density profile (though the results would still be dependent on the
assumption of the age- or metallicity-dependence of the color gradient).

In Table \ref{tab:struct} we list the parameters of the best-fit de
Vaucouleurs, King (1966), and power-law profiles, under the three
extreme assumptions about the gradient in the stellar populations
(the power-law profile is assumed to be of the form: $\rho \propto
[1+(r/r_c)^2]^\gamma$). All three parameterizations fit the data well,
with \rms\ residuals between 0.03 and 0.07 mag.

When fitting the model profiles to the uncorrected $B$- and $R$-band
surface brightness profiles, one obtains smaller effective and core
radii in the redder bandpass, consistent with the stellar populations
becoming bluer with increasing radius. When the $B$- and $R$-band
surface brightness profiles are corrected, the resulting structural
parameters from the two bandpasses are essentially identical. Because we
have divided our observed surface brightness profiles by the gradients
in $M/L_B$ and $M/L_R$, the corrected surface density profiles should
more accurately reflect the stellar mass density profile (modulo the
normalization by a constant).

As an aside, there are large differences in the values of $r_e$ obtained
from the uncorrected, age-variation-corrected, and
metallicity-variation-corrected profiles. In the case where the stellar
population gradient of NGC 6166 (and presumably of other giant
ellipticals) is due solely to gradients in the chemical abundances, the
deduced half-mass radius is smaller than the $B$- and $R$-band
half-light radii by 45\% and 30\%, respectively. Such large systematic
errors in the half-mass radii of elliptical galaxies may have a grave
impact on early-type galaxy scaling relations, such as the fundamental
plane \citep{faber87,dd87}, through the strongly correlated errors in
$r_e$ and $\mu_e$ and the modest scatter in the color gradients of
ellipticals and S0s \citep[e.g.,][]{peletier90}. For the work presented
here, we do not utilize the half-light or half-mass radii, so we offer
this short discussion simply as a cautionary note.

Now that we are free from the assumption of a constant stellar
mass-to-light ratio, we can explore to what extent the stellar mass may
or may not be sufficient to support the motions of the stars through the
cD halo.

\begin{figure*}[t]
\epsscale{1.9}
\plotone{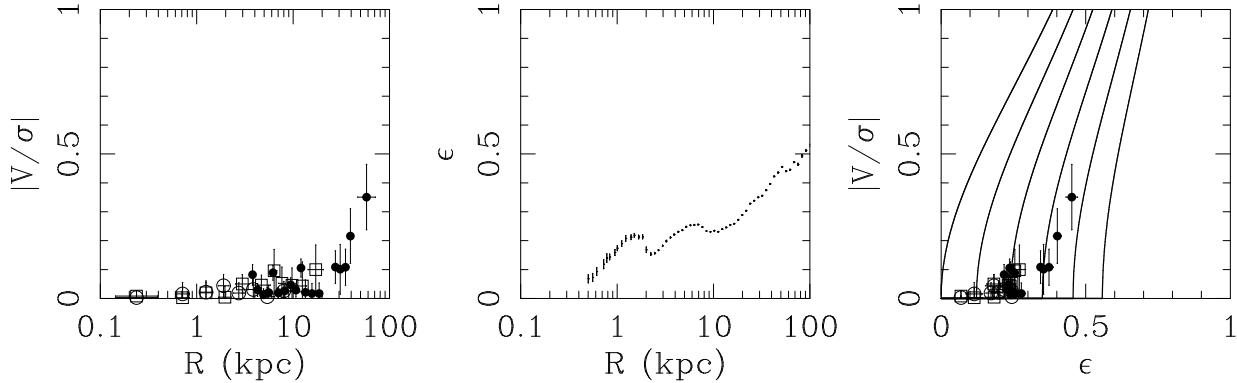}
\epsscale{1.0}
\figcaption[fig8.eps]{(left) The level of rotational support as a function of
projected radius along the major-axis (symbols as in Fig.
\ref{raw_prof}. Most of the stars in NGC 6166 are clearly supported by
random motions with an apparent increase in rotational support outside
of $R= 20$ kpc. (center) The ellipticity of the $B$-band isophotes as a
function of projected radius along the major-axis. The isophotes of NGC
6166 become increasingly elliptical at large radii, consistent with an
increase in rotational flattening. (right) The level of rotational
support as a function isophote ellipticity in the $B$-band. The curves
show the expected correlations between $|V/\sigma|$ and $\epsilon$ for
rotationally supported oblate spheroids with $\delta \equiv
1-\sigma_{zz}/\sigma_{xx}\in \{0,0.1,0.2,0.3,0.4,0.5\}$ observed
edge-on. Using these data we conclude that within a projected radius
$R\simlt 20$ kpc NGC 6166 is not a rotationally flattened oblate
spheroid. Beyond projected radii $R=20$ kpc, the isophotes become more
elliptical and the cD halo appears to become mildly anisotropic and
rotationally flattened. This level of rotational flattening only amounts
to $\simlt 5\%$ of the kinetic energy in the halo and has little effect
on our results.
\label{shape}}
\end{figure*}

%%%%%%%%%%%%%%%%%%%%%%%%%%%%%%%%%%%%%%%%%%%%%%%%%%%%%%%%%%%%%%%%%%%%%%%%

\section{The Major-Axis Kinematics of NGC 6166}
\label{kinematic}

Spectra for the galaxy were extracted from the radial bins defined in
Figure \ref{finder}. Four represenative spectra are shown in the
right-hand panels of Figure \ref{spectra}. In each of the radial bins,
$S/N\ge 20$ per \AA, with the $S/N$ reaching 120 per \AA\ inside 15 kpc.
With such $S/N$ ratios, the data are quite suitable for measuring the
internal motions of the stars in NGC 6166 and its halo out to 60 kpc. To
simplify the derivation of the internal kinematics, we parameterize the
LOSVD as a Gaussian. In the inner regions the high $S/N$ ratios would
normally have allowed us to measure the higher-order moments $h_3$ or
$h_4$ of the LOSVD \citep{vandermarel}. Unfortunately (see below), an
interloping object with a velocity shift of 1300 km/s from NGC 6166
contaminates its spectrum between 4 kpc $\simlt r \simlt $ 12 kpc. As a
result the LOSVDs for both objects were modeled by pure Gaussians, with
both components being fit simultaneously \citep[e.g.,][]{expector}. At
large radii in the cD halo, where accurate determinations of $h_3$ or
$h_4$ would also be useful for constraints on the anisotropy, the $S/N$
ratios are not sufficient. However in other giant ellipticals the
departures of LOSVDs from Gaussian are typically small, even at large
radii \citep[e.g.,][]{kronawitter}.

For each extracted spectrum, we performed a least-squares fit of
broadened template spectra, using stellar template spectra from the
literature \citep{expector} and a more narrow range of stellar spectra
obtained during the observing run. Using the procedures outlined in
Kelson \etal\ (2000), we use night-sky air-glow emission lines and
calibration lamp lines to measure the instrumental broadening. Once the
instrumental broadening had been characterized as a function of
wavelength and position along the CCD, the template spectra from the
literature were broadened to match the instrumental resolution of the
galaxy spectra. While we relied primarily on the templates from the
literature to derive the internal kinematics, the template spectra
obtained during the run were useful to verify the robustness of our
measurements. The velocity dispersions derived using the templates
observed during the run agreed with velocity dispersions derived using
the templates from the literature at the level of 0.5\%.

The quality of the fits to the galaxy spectra can be seen in the
right-hand panels of Figure \ref{spectra}. While the interloper is
clearly visible in two of the examples in the figure, the contamination
was easily removed by the fitting procedures and the residuals of the
fits show no remaining systematic effects. The other two radial bins
shown in the figure show the quality of the velocity dispersion fits at
large projected radii ($17.2\le R\le 20.1$ kpc and $37.3\le R\le 41.3$
kpc).

In the left panel of Figure \ref{raw_prof}, we use filled circles to
show the velocity dispersion profile derived from the deep, primary
long-slit observations on one side of the galaxy (see Figure
\ref{finder}). The open points (circles and squares) show the data from
the shallow central pointing; the open circles are from the same side of
the galaxy as the deep pointing, and the open squares were derived from
the opposite side of the galaxy. The agreement of $\sigma(R)$ from both
sides of the galaxy is excellent and indicates that the interloping
galaxy was successfully removed during the fitting of the deep
spectroscopy. The reduced $\chi^2$ values, shown in the central panel,
show the quality of the velocity dispersion fits.

In the deep spectroscopy of the cD halo there is evidence for a
significant level of rotation beyond $R\simgt 30$ kpc, consistent with
the results of \cite{carter1999} to $R\sim 20$ kpc. The right panel of
Figure \ref{raw_prof} shows the stellar rotation curve and the fraction
of rotational support ($|V/\sigma|$) as a function of projected radius
is shown in the left-hand panel of Figure \ref{shape}.

\begin{figure*}[t]
\epsscale{1.7}
\plotone{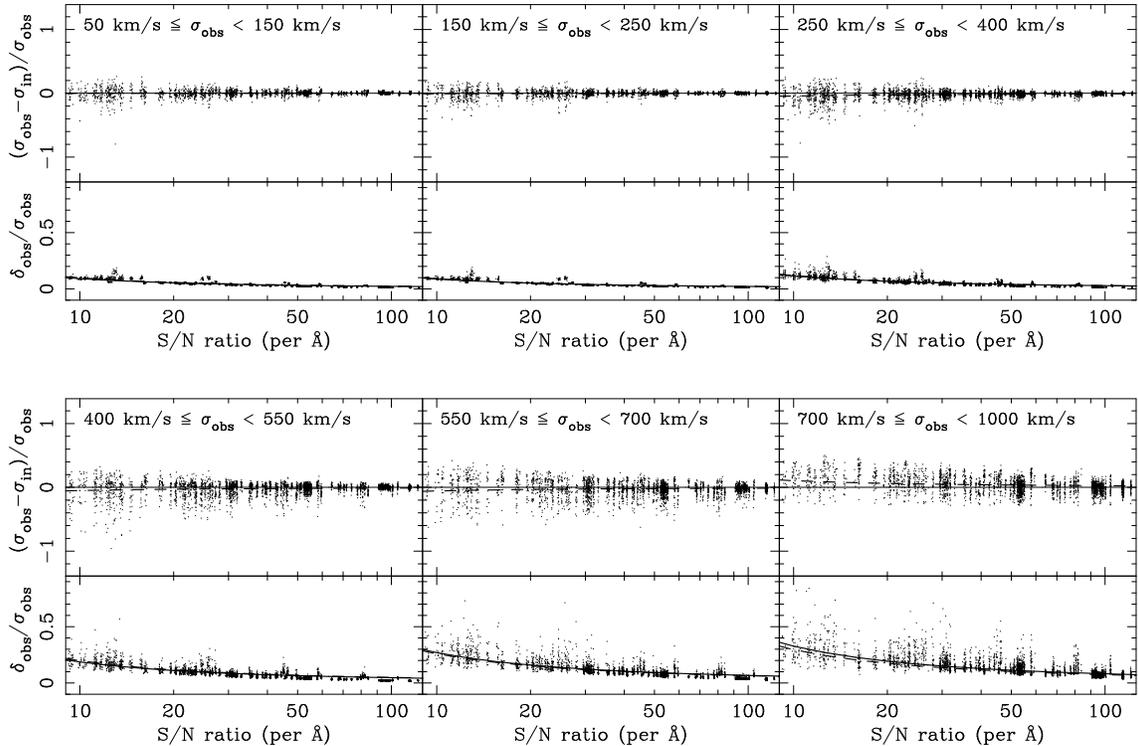}
\epsscale{1.0}
\figcaption[fig9.eps]{Monte Carlo simulations with more than $3\times 10^4$
realizations of our NGC 6166 spectra for which (1) there is no mismatch
between the template and the artificial galaxy spectra; and (2)
surface brightness $I_{in} \ge 0.05 I_{sky}$. The upper panels show the
true fractional errors in $\sigma$ as a function of ``observed''
velocity dispersion, $\sigma_{obs}$. The ``observed'' parameters
$\sigma_{obs}$ and the formal errors, $\delta_{obs}$, were derived by
applying the direct fitting method of Kelson \etal\ (2000) to the
simulated spectra. The velocity dispersions input into the simulated
spectra are denoted by $\sigma_{in}$. Any systematic offsets between the
points in the upper panels from zero would indicate a decrease in our
ability to accurately measure velocity dispersions within a given range
of $\sigma_{obs}$. The dashed lines in the upper panels show the
systematic error in $\sigma_{obs}$ as a function of $S/N$ ratio. These
diagrams indicate that our method can reliably recover velocity
dispersions up to $\sigma_{obs}\simlt 700$ km/s. For larger values of
$\sigma$, there appear to be systematic errors $\simgt +5\%$ for $S/N
\simlt 30$ per \AA. The lower panels show the observed formal errors as
a function of $S/N$ ratio in the given velocity dispersion ranges. The
mean formal error as a function of $S/N$ is shown by the solid line
while the ``true'' formal errors, as derived from the scatter in the
upper diagrams, is shown by a dashed lines in each of the lower panels.
The agreement between the two curves illustrates that our formal errors
are reliable estimates of the random error in the measurement.
\label{montecarlo1}}
\end{figure*}

The error bars shown in the figures represent the formal uncertainties
only. Despite the 18\% formal uncertainty and elevated $\chi^2$ in the
last radial bin, the gradient in $\sigma$ is clearly significant and is
also consistent with the profile of \cite{carter1999} to $R\simlt
20$ kpc. Additional errors, such as those arising from template
mismatch, are estimated at \about 3-5\% by comparing the velocity
dispersions using templates of varying spectral types. Errors in
$\sigma$ due to background subtraction are included in the estimates of
the formal uncertainties \citep[verified by using the $\chi^2$
minimization described in \S 5.2.4 of ][]{expector}. The implications of
the major-axis kinematics for the mass profile of NGC 6166 will be
discussed in \S \ref{models}.

Table \ref{tab:kinprof} lists the major-axis kinematics for NGC 6166.
The columns are as follows: (1) the central radius of the radial bin, in
kpc; (2) the length of the radial bin, in kpc, (3) $\sigma$, the
velocity dispersion, in \kms; (4) $\delta_\sigma$, the formal error in
$\sigma$, in \kms; (5) $V$, the mean line-of-sight radial velocity, in
\kms; (6) $\delta_V$, the formal error in $V$, in \kms; and (7) the
reduced $\chi^2$ of the fit.

%%%%%%%%%%%%%%%%%%%%%%%%%%%%%%%%%%%%%%%%%%%%%%%%%%%%%%%%%%%%%%%%%%%%%%%%

\section{Monte Carlo Simulations of the Accuracy of Velocity Dispersion
Measurements}
\label{montecarlo}

Accurate measurement of velocity dispersions, {\it i.e.}, absorption
line widths, from galaxy spectra requires: (1) that the instrumental
broadening of the galaxy spectrum must be (approximately) less than the
intrinsic Doppler broadening; and (2) that there must be statistically
significant structure on scales comparable to the Doppler broadening. If
either of these requirements is not satisfied, systematic errors may
become important. The great difficulty in measuring the velocity
dispersion profiles of galaxies to large radii, and for cD galaxies in
particular, is that this last requirement may not be satisfied at large
projected radii. In the cD halo, the potential may become dominated by
dark matter and the stars may have similar kinematics to the cluster
galaxies. Furthermore, the surface brightness of the stellar component
of the cD halo is typically quite low (several magnitudes below the dark
sky in the $B$-band). At such low surface brightnesses even a small
error in the background subtraction can introduce unwanted
structure/power into the extracted galaxy spectrum. Because the
representative sky spectrum in Figure \ref{skyspec} has significant
power on the scales relevant for the measurement of absorption line
velocity widths, there exist the possibilities of over-estimated $S/N$
ratios and of excess structure in the spectra at a sufficient level to
bias the deduced LOSVD. In this section we analyze simulations of our
data to determine under what conditions such systematic effects become
important and at what level our measurements of the kinematics in NGC
6166 may no longer be trustworthy.

Because our data span a wide range of measured velocity dispersions over
a broad range of surface brightnesses, we constructed 4680 artificial
galaxy spectra with intrinsic broadening of $\sigma_{in}\in
\{100,200,300,400,500,600,700,800,900\}$ km/s. We used the 10 Gyr-old,
solar metallicity, blue SED from Vazdekis (1999) as the underlying
spectrum of the artificial galaxy. These spectra reproduce the
characteristics of the NGC 6166 data, assuming a range of surface
brightnesses, $I_{in} \in \{3.0,1.0,0.5,0.1,0.05\} \times I_{sky}$, and
sky subtraction errors, $E_{in} \in \{\pm 0.02, \pm 0.01, \pm 0.005, \pm
0.001, 0.0\} \times I_{sky}$. The artificial spectra were given a range
of $S/N$ ratios equivalent to summing several CCD rows as in \S
\ref{extraction} ($10 \le N_{rows} \le 1000$, depending on the desired
$S/N$ ratios).

With the many variations described in the previous paragraph, $3\times
10^4$ artificial spectra were created with $5 \simlt S/N \simlt 200$ per
\AA. The key results of fitting these simulated galaxy spectra with a
perfect template are shown in Figure \ref{montecarlo1}. In the upper
panels, for several ranges of ``observed'' velocity dispersions
($\sigma_{obs}$), we show the fractional error in $\sigma$,
$(\sigma_{obs}-\sigma_{in}/\sigma_{obs}$, as a function of $S/N$ ratio.
If the data points cluster about the solid lines, then our methods are
free from any systematic errors (template mismatch is not explored in
this figure). The dashed lines show the best-fit curve of
$(\sigma_{obs}-\sigma_{in}/\sigma_{obs} \propto [\log(S/N)]^{-2}$. For
$\sigma_{obs}\simlt 700$ km/s, the measured velocity dispersions are
free from systematic errors for $S/N \simgt 10$ per \AA. Figure
\ref{montecarlo1} suggests the presence of systematic errors when
$\sigma_{obs} \simgt 700$ km/s and $S/N \simlt 30$ per \AA, but this is
an artifact of the restriction $\sigma_{in} \le 900$ km/s (in other
words, $\sigma_{obs}=800$ km/s with an error of $-20\%$ does not appear
because there were no artificial spectra with $\sigma_{in}=960$ km/s).
Based on the upper panels we conclude that the velocity dispersion
profile derived in \S \ref{kinematic} is free from systematic errors
(other than template mismatch, discussed below).

The lower panels of Figure \ref{montecarlo1} show the formal errors,
$\delta_{obs}$, in $\sigma_{obs}$ as a function of $S/N$ ratio. The
solid lines show the best-fit functions of the form
$\delta_{obs}/\sigma_{obs} \propto [\log (S/N)]^{-2}$ and illustrate
the correlation of the estimated formal errors in $\sigma$ with $S/N$
ratio. These estimates for the formal errors can be empirically verified
using the scatter in the upper diagrams. The best-fit functions (using
the same form), which represent the correlations of the true random
errors with $S/N$ ratio, are shown by the dashed lines in the lower
panels. The agreement between the two curves is excellent and indicates
that we have satisfactorily estimated the random errors in $\sigma$.

The simulations were also used to explore whether fitting the Ca H \& K
features improves one's ability to resolve large values of the Doppler
broadening. The 4000\AA\ break and the H \& K lines are broad features
with presumably sufficient power on the scales required to
accurately measure large velocity dispersions ($\sigma \simgt 500$
km/s). However, incorporating that region of the spectrum does not
improve the measurement of $\sigma$ because the $S/N$ ratios are much
lower than those in the region around the G band, a more useful feature
because of the confluence of the higher fluxes and the intrinsic 2000
km/s width of the feature.

We also used the simulations to explore the effects of template
mismatch. By fitting the artificial galaxy spectra with spectra of
individual late-type stars, we found small systematic errors on the
order of $\simlt 5\%$. These systematic errors are not correlated with
$S/N$ ratio, and if one adopts the template that produces the lowest
$\chi^2$, the error is minimized. Such systematic uncertainties, at a
level of a few percent, are likely to be present in our velocity
dispersion profile for NGC 6166.

The simulations shown in Figure \ref{montecarlo1} also point to an
additional effect that will impact future studies: for a given desired
fractional error in $\sigma$, large velocity dispersions require data
with higher $S/N$ ratios. For example, measuring an internal velocity
dispersion of $\sigma=900$ km/s with an accuracy of $\pm 10\%$ requires
spectra with $S/N=50$ per \AA. An accuracy of $\pm 15\%$ requires
$S/N=30$ per \AA. Fortunately we did not find such large velocity
dispersions in \S \ref{kinematic}, even at a projected radius of 60 kpc.
In the future, however, when the observer pushes towards larger radii
and/or lower surface brightnesses in other targets, such levels of
Doppler broadening may occur, and acquiring data with such high $S/N$
ratios will be necessary to produce credible measurements of the
internal kinematics. This requirement may be difficult to meet with
current 8m-10m telescopes (even using the sky-subtraction technique of
\citep{tonry2}.

In summary we conclude that the velocity dispersion profile derived in
\S \ref{kinematic} is free from systematic errors at the level of $\pm
3\%$. Most of the systematic error arises from a mismatch of the
template star with the galaxy.

%%%%%%%%%%%%%%%%%%%%%%%%%%%%%%%%%%%%%%%%%%%%%%%%%%%%%%%%%%%%%%%%%%%%%%

\begin{figure*}[t]
\epsscale{1.3}
\plotone{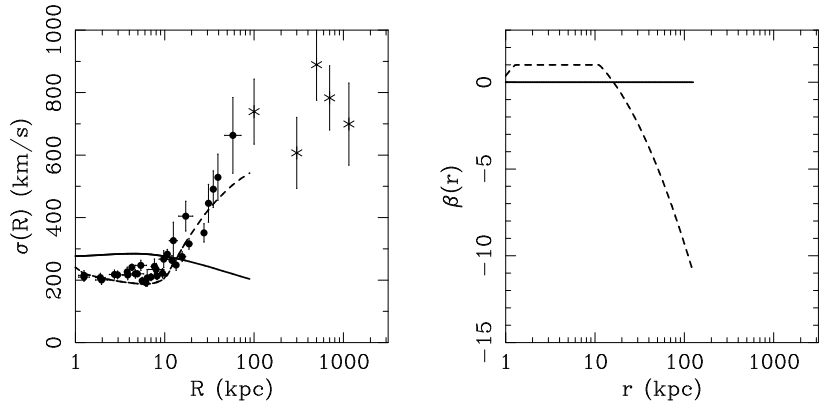}
\epsscale{1.0}
\figcaption[fig10.eps]{The observed velocity dispersion profile of NGC
6166 and its halo from integrated starlight (filled circles) and Abell
2199 from cluster member kinematics (crosses). The solid line is the
projected line-of-sight velocity dispersion profile of the best-fit
isotropic single-component stellar mass model in which a power-law
stellar density profile is used. The dashed line illustrates that the
rising velocity dispersion profile of NGC 6166 can be well fit by an
anisotropic model, while the right-hand panel shows the anisotropy used
in the two models. Such high levels of anisotropy are not seen in other
giant ellipticals and BCGs (typically $-0.5 \simlt \beta \simlt 0.5$;
\citep{saglia,kronawitter,gerhard}, and result in mass models that
cannot support the velocity dispersion of the cluster unless the
galaxies also follow such strongly tangentially anisotropic orbits.
\label{stellarmod}}
\end{figure*}

\section{Simple Spherically Symmetric Models}
\label{models}

Under dynamical equilibrium, the observed velocity dispersion profile is
a constraint on the density profile \citep{jeans}. In principle one can
invert the observed kinematics to directly infer the underlying
distribution of mass \citep{bm82} or use orbital-superposition
approaches to build realistic mass models
\citep{schwarzschild,richstone,rix}. However, the construction of
anisotropic mass models requires measurements of the Gauss-Hermite
moments $h_3$ and $h_4$, and we did not achieve sufficiently high $S/N$
ratios to measure these higher order moments of the LOSVD in the outer
parts of NGC 6166 \citep[$\simgt 40$ per \AA\ is required;
][]{vandermarel}, and our measurements in the inner parts were corrupted
by the interloper (\S \ref{kinematic}).

With kinematics from only one position angle, and without the high-order
moments of the LOSVD, several assumptions must be made before deriving
meaningful constraints on the mass profile. For the modeling of NGC 6166
and Abell 2199, we assume spherical symmetry and zero net rotation. Even
though the isophotes of NGC 6166 have ellipticities ranging from 25\% at
$R=5$ kpc to to 50\% at $R=75$ kpc (Fig. \ref{shape},  and the cD halo
appears to be rotationally flattened, this rotation accounts for less
than $\simlt 5\%$ of the kinetic energy in the cD halo. Because neither
the bulk of the cD, nor the cluster member kinematics, is supported by
rotation, the mild rotation in the outskirts of the cD has little impact
on the modeling. We also note that the ellipticity of the isophotes does
not necessarily invalidate our assumption of spherical symmetry --- the
total gravitational potential is most likely close to spherical given
the morphology of the cluster.

Thus this section is devoted to fitting simple spherically symmetric 1-
and 2-component mass models to the velocity dispersion profile of NGC
6166. The reduction of the Jeans equations to spherically symmetry
greatly simplifies the computation of the projected line-of-sight
velocity dispersion profiles, $\sigma(R)$, for the model mass profiles
we explore below.

For non-rotating spherical systems the Jean's equations
(Equations 4-55, 4-57a, and 4-60 of \citep{bt87}
are:\footnote{The integrals are evaluated numerically using the ODEPACK
and QUADPACK ({\tt http://www.netlib.org/}) routines from within Python
({\tt http://www.python.org/}) scripts.}
\begin{eqnarray}
{G M(r) \over r} &=& -\overline{v_r^2}\bigl(
{d \ln \nu \over d\ln r} +
{d \ln \overline{v_r^2} \over d\ln r} + 2\beta(r)\bigr)\\
I(R)\label{eq:surf}
&=& 2\int_R^\infty {\nu r dr \over \sqrt{r^2-R^2}}\\
I(R)\sigma^2(R)\label{eq:sig}
&=&
2\int_R^\infty
\bigl(1- \beta(r) {R^2\over r^2}\bigr)
{\nu \overline{v_r^2} r dr \over \sqrt{r^2-R^2}},
\end{eqnarray}
where $\beta(r)\equiv 1- {\overline{v_\theta^2}/\overline{v_r^2}}$,
$\nu$ is the stellar density at $r$ as constrained by the surface
photometry, and $M(r)$ is the total gravitational mass enclosed within
radius $r$. The gravitational mass, $M(r)$, includes the stellar mass
and relies on assumptions about the radial gradient in the stellar $M/L$
ratio. With the radial gradients in $M/L_B$ suggested in \S
\ref{stellarmass}, the stellar density decreases more quickly than in
the case of constant $M/L_B$. However, in the case of a decreasing
$M/L_B$ gradient, the light-weigted integrals in Equations \ref{eq:surf}
and \ref{eq:sig} are weighted towards larger radii, where the stars are
increasingly more luminous per unit mass.

For each parameterized model, we perform a search for the parameters
that minimize the square of the residuals between the model and observed
velocity dispersion profiles. We choose to uniformly weight the data
because weighting by the inverse of the formal errors would cause the
minimization algorithm to find the parameters that best fit the inner
regions of the galaxy. Furthermore, we exclude the inner $R=2''\equiv
1.1$ kpc from the fit because: (1) this region is heavily affected by
the seeing (recall: $0\Sec 8$ FWHM); and (2) the spectrum in the center
shows evidence for nuclear activity \citep[also noted by][]{fisher} and it
is likely that a massive black hole dominates the kinematics in the very
center, given the correlation between galaxy and central black hole
masses \citep{karl}.

Given the simplicity of the models, we attempt no derivation of
uncertainties in the parameters of each model as their meaning could
only be as valid as any given model is an appropriate representation of
the mass distribution, or as valid as any of the underlying assumptions.
Instead we use the $\chi^2$ minima to assess whether given mass models
are consistent with the data at a given confidence level ($\ge 95\%$).

%%%%%%%%%%%%%%%%%%%%%%%%%%%%%%%%%%%%%%%%%%%%%%%%%%%%%%%%%%%%%%%%%%%%%%%%

\begin{figure*}[t]
\epsscale{0.6}
\plotone{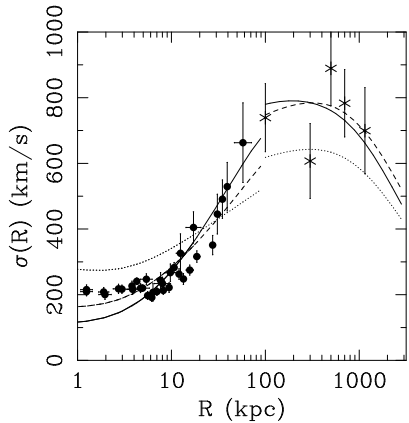}
\epsscale{1.0}
\figcaption[fig11.eps]{The observed velocity dispersion of NGC 6166 and its halo
from integrated starlight (filled circles) and Abell 2199 from cluster
member kinematics (crosses) as a function of projected radius with the
single-component dark matter models overlayed. The solid, dashed, and
dotted lines show the best-fit $\alpha=0,1,1.5$ generalized-NFW density
profiles, respectively.
\label{darkmod}}
\end{figure*}

\subsection{Single-Component Mass Models}

In this section we fit the data with models containing a single mass
component. Two single-component mass models are considered here: (1) a
distribution of stars alone; and (2) a distribution of dark matter
alone, in which the stars are effectively massless tracers of the dark
potential (the stellar $M/L$ is identically zero). While fitting
single-component mass models is unlikely to provide physically
meaningful characterizations of the distribution of mass in the core of
Abell 2199 given the presence of both stars and dark matter, these
models provide natural starting points for illustrating to what extent
additional components might be required. After a brief discussion of the
deficiencies of the single-component models, two-component mass models
will be discussed in \S \ref{two}

%%%%%%%%%%%%%%%%%%%%%%%%%%%%%%%%%%%%%%%%%%%%%%%%%%%%%%%%%%%%%%%%%%%%%%%%

\subsubsection{Single Component Stellar Model}
\label{stellar}

For this exercise, the observed velocity dispersion profile of NGC 6166
is fit using a mass density model defined by the fit of a King (1966)
model to the surface brightness profile (see \S \ref{stellarmass},
under the assumption of a constant $M/L_*$ ratio. The structural
parameters for this model are listed in Table \ref{tab:struct}, leaving
the overall normalization of $M/L_*$ and the anisotropy as the only free
parameters. For this model we explore two variants: the isotropic case
and the case in which the anisotropy, $\beta(r)$, has been adjusted to
provide a reasonable match to the dispersion profile.

We show these models in Figure \ref{stellarmod}(a), using the solid line
to denote the best-fit isotropic model and the dashed line to illustrate
the best-fit anisotropic model. Figure \ref{stellarmod}(b) shows the
anisotropy required to reproduce the observed velocity dispersion
profile. These models illustrate that the stellar mass alone cannot
support the rising $\sigma(R)$ profile in the cD halo, unless the
stellar orbits become significantly (and unreasonably) tangentially
anisotropic \citep[also see][]{tonry}. Such levels of anisotropy are
inconsistent with what has been observed at 1-2 $r_e$ in other brightest
cluster galaxies \citep[typically $-0.5 \simlt \beta \simlt 0.5$;][]
{saglia,kronawitter,gerhard}. 

The purely stellar models imply a total mass of $M_*= 4\times 10^{12}
M_\odot$. This mass is insufficient to support the observed velocity
dispersion of the cluster galaxies $\sigma_{\rm A2199}=775\pm 50$ \kms,
unless the galaxy kinematics become even more strongly tangentially
anisotropic (essentially following an extrapolation of the dashed line
in the figure). Figure \ref{stellarmod}(a) shows the velocity dispersion
profile of the cluster derived from the 127 member galaxies within $R\le
1.5$ Mpc (obtained from NED). Given that the orbits of cluster galaxies
are typically close to isotropic \citep{priya,vandermarel2}, it is
unlikely that the galaxy orbits in Abell 2199 are as anisotropic as
would be required with the purely stellar mass model. Even if $M/L_*$ is
not a constant, under the constraints of \S \ref{stellarmass}, $M(R)$
converges even more rapidly. As a result, the outskirts of NGC 6166
would be required to be more strongly anisotropic than in the case of a
constant $M/L_*$.

We consider the single-component stellar models unfeasible because both
the stellar and galactic orbits are required to be unreasonably
tangentially anisotropic in order to support the large observed velocity
dispersions.

%%%%%%%%%%%%%%%%%%%%%%%%%%%%%%%%%%%%%%%%%%%%%%%%%%%%%%%%%%%%%%%%%%%%%%%%

\subsubsection{Single Component Dark Matter Model}
\label{single:dark}

We now model the data using a single-component dark mass profile, in
which the stars of NGC 6166 are massless tracers of the potential. In
order to compute the projected line-of-sight velocity dispersion profile
for NGC 6166, we use the stellar luminosity density profile (the
projected velocity dispersion for the galaxy is a {\it light\/}-weighted
moment of the LOSVD). At larger radii, where the kinematics of cluster
galaxies are sampled, we assume that the galaxy number density traces
the density of the dark matter model. This assumption is valid because
van der Marel \etal\ (2000) found that galaxy number density profiles of
the CNOC1 clusters traced the mean dark matter density profile well.

In Figure \ref{darkmod} we show the best-fit models using
generalized-NFW density profiles with $\alpha\in\{0,1,1.5\}$. Because
the NGC 6166 data-points have smaller relative uncertainties, the cD
data carry more weight than the cluster data, and the extremely cuspy
$\alpha=1.5$ model simply cannot be adjusted to fit both the inner
regions of the cD while maintaining the overall normalization of the
total mass as constrained by the cluster data. Even for the softer
halos, a single component model cannot fit the galaxy and the cluster
velocity dispersion profile simultaneously. The soft $\alpha\le 1$ model
fits the rise in $\sigma(R)$ and the cluster data quite well, but leaves
a hole in the inner region where the non-zero stellar mass of the cD
resides.

The concentration parameter, $c=R_{200}/r_s$, is left as a free
parameter in the fit. The best-fit $\alpha=1$ profile has $c=4$ and
$R_{200}=1.6$ Mpc. Bullock \etal\ (2001) discussed the scatter in
concentration parameters as a function of halo mass, and while they do
not address halos as massive as ours ($M_{200}=5\times 10^{14}M_\odot$),
an extrapolation of their Figure 4 indicates that $c=4$ would fall
within the $\pm 1\sigma$ range of concentration parameters at this halo
mass.

Figure \ref{darkmod} also illustrates an important physical effect in
the mass modeling: that if the density profile of the cD halo does not
trace the number density profile of the galaxies in the cluster, then
the velocity dispersion profiles of the two components will not be
identical at a given projected radius. \citep[also see][]{priya}. This
effect appears as a discontinuity between the model velocity dispersion
profiles of the cD and the cluster. As the minimization algorithm
adjusts the shape of the dark matter density profile, the inferred
number density profile of cluster galaxies may not have the same shape
as the stellar density profile in the outskirts of the cD. As a result,
the best-fit dark matter model may imply a distribution of galaxies that
traces the potential differently than the stars in cD halo do (at a
fixed projected radius). This difference results in a velocity
dispersion profile for the cD that may not seamlessly join the velocity
dispersion profile of the cluster. Under different assumptions (forcing
the stars in the cD halo to have originated from the current
distribution of cluster galaxies), the velocity dispersion profiles of
the cD and the cluster can be made a single, continuous function.
Perhaps a future analysis of the color and absorption line strength
gradients will help to constrain nature of the relationship between the
stars in the cD halo and the cluster galaxies.

In this section we restricted ourselves to isotropic dark matter models,
a reasonable assumption because the CNOC1 clusters appear to be
isotropic \citep{vandermarel2}, and also because X-ray observations of
Abell 2199 constrain the anisotropy in the dark matter velocity
distributions to $-0.5 \simlt \beta \simlt 0.5$ over a wide range of
cluster radii \citep{mahdavi}. The conclusions drawn from the models in
this section are insensitive to these levels of anisotropy.

%%%%%%%%%%%%%%%%%%%%%%%%%%%%%%%%%%%%%%%%%%%%%%%%%%%%%%%%%%%%%%%%%%%%%%%%

\begin{figure*}[t]
\epsscale{1.3}
\plotone{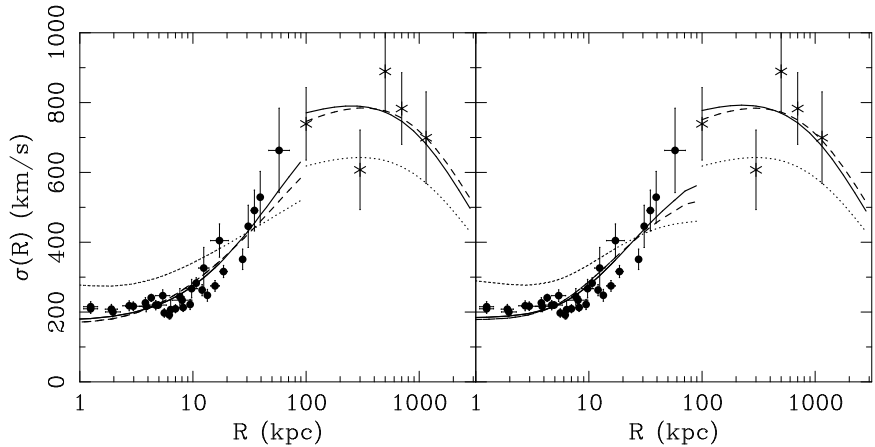}
\epsscale{1.0}
\figcaption[fig12.eps]{The observed velocity dispersion of NGC 6166 and its halo
from integrated starlight (filled circles) and Abell 2199 from cluster
member kinematics (crosses) as a function of projected radius,
with simple two-component models overlayed in which (a) a power-law
density profile or (b) King (1966) model is used to represent the
stellar mass. These models assume a constant $B$-band stellar $M/L_*$
ratio and the solid, dashed, and dotted lines indicate the models with
the generalized-NFW dark matter halos for which $\alpha=0,1,1.5$,
respectively. The models with $\alpha=0$ have stellar $M/L$ ratios
$M/L_B=9$, consistent with the $M/L_*$ ratios seen in other giant
ellipticals \citep{gerhard}. With $\alpha=1$, $M/L_B=1.2$ and $0.2$
when the two stellar mass density profiles are used, respectively. Using
$\alpha=1.5$ forces $M/L_B=0$, and these models still have far too much
dark matter in the core to adequately match the observed velocity
dispersion profile.
\label{stars+nfw}}
\end{figure*}

\subsection{Two-Component Mass Models: Stars $+$ Dark Matter}
\label{two}

The previous section showed that one-component isotropic mass models do
not satisfactorily reproduce the projected line-of-sight velocity
dispersion profile of NGC 6166. With this discrepancy in mind, we take
the dark models and add a component that physically represents the
stellar mass of the cD. The stellar mass profile is directly constrained
by the surface photometry, with the overall normalization (the stellar
$M/L$ ratio) to be constrained by $\sigma(R)$ in the inner regions of
the galaxy.

In this section we find the best-fit two-component isotropic mass models
\citep[see earlier references to][regarding
isotropy]{saglia,kronawitter,vandermarel2,mahdavi}. The two mass
components are expected to represent the stellar mass of the cD and the
dark matter distribution of the cluster. Because the surface brightness
profiles of giant ellipticals are well-fit by power-law and King (1966)
density models \citep{king78,lauer,graham}, we explore models in which
the stellar component is modeled either by a power-law mass density
profile or by a King model density profile, with structural parameters
given in Table \ref{tab:struct}. The dark matter distribution of the
cluster is parameterized by the generalized-NFW halo profile.

In Figure \ref{stars+nfw}(a), we show the best-fit models for the case
where a power-law density profile of constant $M/L_*$ ratio is used to
parameterize the stellar mass of the cD and where generalized-NFW
density profiles with $\alpha\in\{0,1,1.5\}$ are used to parameterize
the dark matter halo. Figure \ref{stars+nfw}(b) uses a King (1966) model
for the stellar mass profile, and the dark matter halo density profiles
are parameterized as in \ref{stars+nfw}(a). When fitting models with an
$\alpha=1$ halo, we find concentration parameters of $c=3.5$ and $4.5$
depending on whether the stellar mass is modeled as a power-law or King
model, respectively. As above, these $c$ values are consistent with the
$\pm 1\sigma$ range for the concentration given by Bullock \etal\
(2001).

Of these two-component models, the dark matter halo profiles with
$\alpha=0$ fit best, yielding a $B$-band $M/L_*$ ratio of $M/L_B=9$ for
either parameterization of the stellar density profile. If $\alpha=1$,
the power-law density and King parameterizations have $M/L_B=1.2$ and
$0.2$, respectively. Using a dark matter halo with $\alpha=1.5$ cannot
reproduce the observed velocity dispersion profile because the halo
density profile has too much dark matter in the core (thus forcing
$M/L_*=0$). The stellar $M/L_B$ ratio derived using $\alpha=0$ is
consistent with what has been derived for other giant ellipticals and
BCGs 
\citep[$8\simlt M/L_B \simlt 12$ for $M\simgt 10^{11} M_\odot$;][]
{gerhard}. The $\alpha=1$ model is unlikely because the deduced
$M/L_B$ is inconsistent with the $B-R$ color of the galaxy (for
realistic initial mass functions, an extrapolation of the \citep{vaz96}
models to $M/L_B=1$ and $B-R=1.5$ implies a mean stellar population age
of 24 Myr and [Fe/H] = +2.4).

Allowing for the gradient in $M/L_B$, as constrained in \S
\ref{stellarmass}, does not solve the problem. As stated earlier, the
color gradient implies that $M(R)$ converges rapidly for the stellar
component. While the implication is that there is less stellar mass at
large radii, the normalization of the stellar mass density profile is
determined by the velocity dispersions within $R\simlt 10$ kpc. Together
these effects produce a defficiency of stars at large radii, but there
is no net impact on the model velocity dispersion profiles because the
dark matter halo is adjusted during the $\chi^2$ minimization to
maintain the total mass profile of the system. Furthermore, if the
stellar mass profile converges very rapidly, as is the case where the
color gradient arises from a gradient in the mean stellar population
ages, more dark matter is required to support the motions of the stars
in the outskirts of the galaxy. This need for additional dark matter at
such radii is found by shrinking the core radius of the dark matter
halo, but not by making the core cuspier (see earlier discussion). As a
consequence, our conclusions are actually insensitive to the assumption
of a constant $M/L_*$ ratio. Only in systems without a massive,
compensating component of dark matter at large radii, will the gradient
in stellar populations have a significant impact on the inferred total
mass profile.

%%%%%%%%%%%%%%%%%%%%%%%%%%%%%%%%%%%%%%%%%%%%%%%%%%%%%%%%%%%%%%%%%%%%%%%%

\subsection{Additional Points}

There are some additional limits to our data that should be noted. For
example, the two-component mass models that employ the power-law density
profile for the stellar mass produce a steeper $\sigma(R)$ gradient
beyond $R=20$ kpc than the equivalent models that use King stellar mass
profiles. This difference arises because the power-law density profile
has infinite extent, with the density falling as $\rho\propto
r^{2.5-3}$, and the King model has finite extent. The surface photometry
is not well-fit by an isothermal sphere and, as a result, the King model
fit is truncated at a finite radius. In the integration to large radii
there will simply be more light at larger radii with a power-law stellar
density profile and thus a steeper gradient in $\sigma(R)$.

We also investigated whether a King model was a better parameterization
for dark matter density profile. Both the King model and $\alpha=0$
generalized-NFW halo have similarly shaped density profiles in the inner
regions, but they transition, respectively, to $\rho \propto r^{-2}$ and
$\rho \propto r^{-3}$ at large radii. Unfortunately the current data do
not extend beyond 1.5 Mpc, and we are therefore unable to determine
whether the density profile of the dark matter halo declines either as
$\rho\propto r^{-2}$ or as $\rho\propto r^{-3}$. A better determination
of the cluster density profile at large radii awaits membership surveys
that extend beyond $R_{200}$ \citep[e.g., as in][]{rines}. Such data would
not only provide greater leverage on the shape of the mass profile at
large cluster radii, but such membership information would allow for a
test of the assumption that the galaxy number density traces the dark
matter density.

%%%%%%%%%%%%%%%%%%%%%%%%%%%%%%%%%%%%%%%%%%%%%%%%%%%%%%%%%%%%

%\subsection{Comparisons with Results from Other Techniques}
%%\label{comparison}

\subsection{Comparison with X-Ray Masses}

At large radii, the inferred mass of the system is insensitive to the
model adopted for the stellar density profile and mildly sensitive to
the adopted dark matter density profile. With the $\alpha=0$
generalized-NFW halo, we obtain $M(<0.5h^{-1}\, {\rm Mpc}) = 1.7 \times
10^{14}h^{-1}) M_\odot$ ($h=H_0/100$) when adopting either the power-law or
King model stellar density profile (and $M(<0.5h^{-1}\, {\rm Mpc}) = 1.5
\times 10^{14}h^{-1} M_\odot$ with $\alpha=1$). These mass estimates are
consistent with the mass derived from ASCA and ROSAT X-ray data:
$M(<0.5h^{-1}\, {\rm Mpc}) = (1.45\pm 0.15) \times 10^{14} h^{-1}
M_\odot$ \citep{markevitch}.

For radii within $\sim 0.1h^{-1}\, {\rm Mpc}$, most X-ray mass
determinations have been hampered by the relatively poor spatial
resolution of the instruments and by what may be a central cooling flow
\citep{sidd,markevitch}. As a result, there has been some variation in
the published mass estimates: Markevitch \etal\ found $M(<0.1h^{-1}\,
{\rm Mpc}) = (0.33\pm 0.06) \times 10^{14} h^{-1} M_\odot$, whereas
Siddiqui \etal\ (1998), who use only the ROSAT PSPC data, found
$M(<0.1h^{-1}\, {\rm Mpc}) \approx 0.22 \times 10^{14} h^{-1} M_\odot$
(from their Figure 15). The difference between Markevitch \etal\ and
Siddiqui \etal\ stems from the need to extrapolate the mass profile into
the region of the cooling flow: Markevitch \etal\ (1999) modeled the
X-ray temperature profile of the cluster using a polytrope and Siddiqui
\etal\ (1998) modeled the cluster using a King profile. Because
polytropes are, by definition, cuspier than the isothermal cores of King
models, Siddiqui \etal\ (1998) infer a shallower mass profile. Using
X-ray data from Chandra, with much higher spatial and spectral
resolution, Johnstone \etal\ (2002) suggest that the NFW profile
provides a marginally better description of the {\it total\/}
gravitational mass profile than $\alpha=1.5$ or a non-singular
isothermal sphere (also see \S \ref{single:dark}). Using their best-fit
NFW model, those authors infer $M(<0.1h^{-1}\, {\rm Mpc}) \approx 0.15
\times 10^{14} h^{-1} M_\odot$. Despite the uncertainties in the X-ray
mass determinations, their estimates are in reasonable agreement with
the values yielded by our two-component models: $M(<0.1h^{-1}\, {\rm
Mpc}) = 0.17-0.19 \times 10^{14} h^{-1} M_\odot$.

\begin{figure*}[t]
\epsscale{1.6}
\plotone{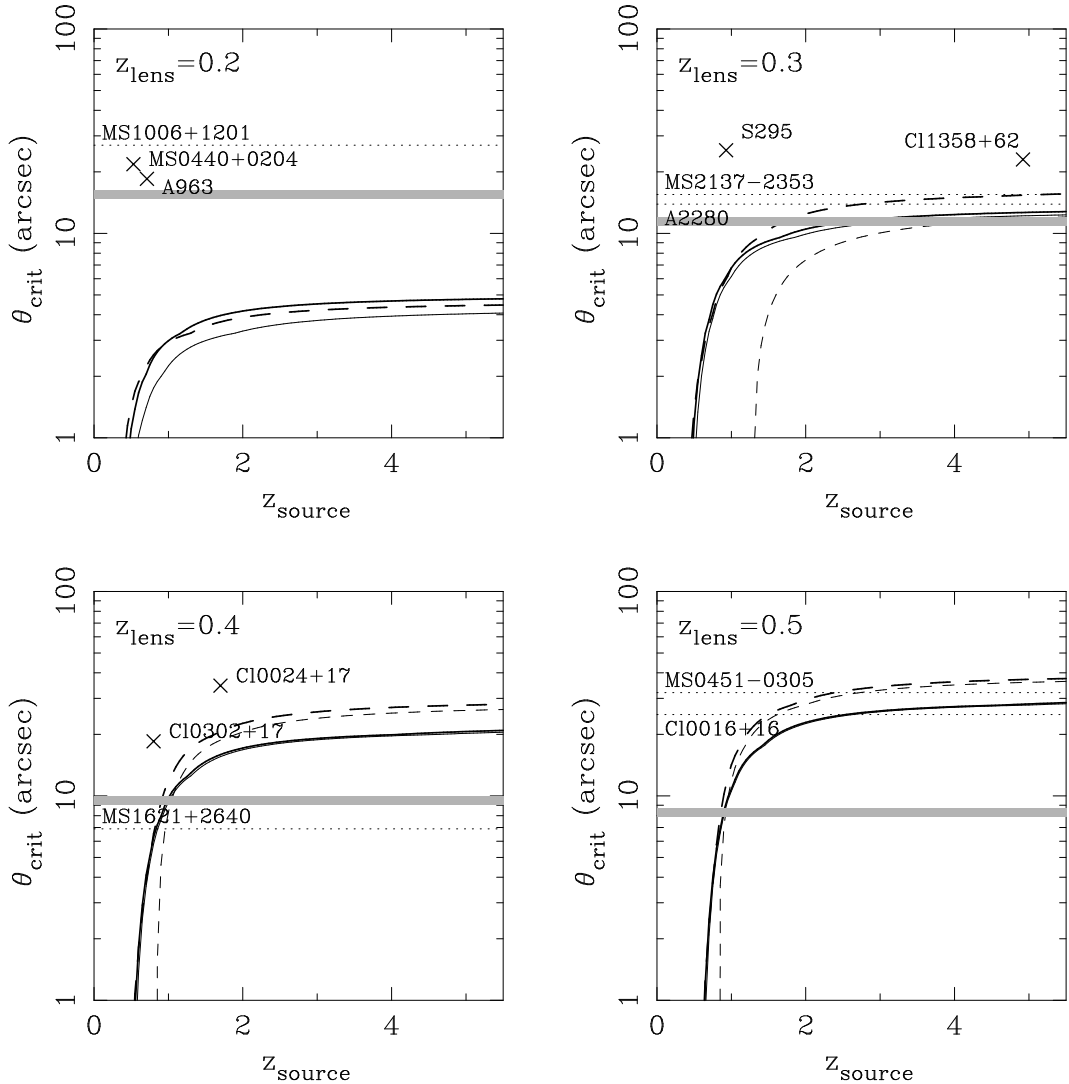}
\epsscale{1.0}
\figcaption[fig13.eps]{The projected radii for the tangential critical
curves as a function of source redshift for several possible lens
redshifts for our two-component mass models. We show the best-fit
two-component mass models in which the stellar density profile is
parameterized by a power-law and the dark matter halo is parameterized
either by $\alpha=0$ (thick dashed) or by $\alpha=1$ (thick solid). The
thin dashed and thin solid lines show the critical radii derived from
the dark matter halos of those two-component models alone, ignoring the
mass of the cD. Using the compilation of tangential arcs in Williams
\etal\ (1999), we show example arcs seen in intermediate redshift
clusters, using crosses when the source redshift is known and horizontal
dotted lines when the source redshift is unknown. The horizontal gray
bars indicate the effective radius NGC 6166 would have at each
$z_{lens}$. If Abell 2199 were redshifted to cosmological distances, the
cluster's mass profile would produce arcs similar to those seen in
intermediate redshift clusters. For the case of the $\alpha=0$ dark
matter halo, the stellar mass serves as a large boost to the lensing
signal. In the two-component mass model with $\alpha=1$, the stellar
mass-to-light ratio is too low for the addition of the stellar mass to
significantly affect the critical radii. Note that at $z_{lens}=0.2$,
the surface brightness of the cD would likely mask any strong arcs.
Furthermore, because the greatest probability for lensing occurs when
$z_{source}= 2z_{lens}$, the predicted $\theta_{crit}$ would also be
difficult to observe with the limited spatial resolution of ground-based
imaging.
\label{lensing}}
\end{figure*}

\subsection{Reconciling Soft Halo Profiles with Strong Gravitational
Lensing}

Given its low redshift, there are currently no gravitational lensing
data for Abell 2199. However, it is possible to ask whether the strong
lensing characteristics of our our two-component mass models are
consistent with the properties of known strong lenses at intermediate
redshifts. Figure \ref{lensing} shows a plot of the tangential critical
radii of our two-component mass models when placed at the
redshifts $z_{lens}\in \{0.2,0.3,0.4,0.5\}$. We restrict the figure to
only include the two-component models with dark matter halo density
profiles with $\alpha=0$ and 1, because these models are able to
reproduce the internal kinematics of NGC 6166 and the surrounding
cluster. We mark the radii for tangential arcs in several intermediate
redshift clusters, using crosses when the source redshift is known, and
horizontal dotted lines when the source redshift is unknown. The
horizontal shaded bar marks the $R$-band effective radius (see Table
\ref{tab:struct} that NGC 6166 would have at $z_{lens}$. At
$z_{lens}=0.2$, the strong lensing would be difficult to observe because
the resulting arcs would be deep inside the image of the cD. Beyond
$z_{lens} \simgt 0.3$, the tangential critical radii lie well outside
$r_e$ and any strong lensing would be observable. The radii of arcs in
clusters at $z_{lens}\simgt 0.3$ are in rough concordance with our model
predictions.

As an additional note, the model with an $\alpha=0$ NFW halo yields
larger critical radii than the $\alpha=1$ NFW halo. This effect occurs
because a more massive central galaxy is needed to reproduce the
observed velocity dispersions within the central 10 kpc, and this extra
structure in the mass profile gives the lensing signal a boost. While
the current modeling has neglected departures from spherical symmetry
and substructure, such effects serve to increase the predicted critical
radii and improve the agreement \citep{bartelmann}. Furthermore, if the
dark matter halo density profile transitions at large radii to $\rho
\sim r^{-2}$ instead of $\rho \sim r^{-3}$, the excess surface mass
densities can increase the predicted critical radii by factors of a few.

In Figure \ref{lensing}, the predicted critical radii increase with
$z_{lens}$ for a given $z_{source}$. However, there appears to be no
significant correlation of the observed arc radii with $z_{lens}$. The
natural distribution of $\theta_{crit}$ may not be as narrow as
indicated by the observed arc radii, but may be truncated at low
$\theta_{crit}$ by the presence of large brightest cluster galaxies and
by the limited spatial resolution of ground based imaging. These effects
are likely to be the cause of the poor agreement between the predicted
critical radii of our mass models at $z_{lens}=0.2$ and the arc radii
seen in $z\approx 0.2$ clusters.

The strong gravitational lensing by MS2137--2353 has been used to
constrain the mass profile of the cluster \citep{jordi95}, and to
predict a rising line-of-sight velocity dispersion profile for the
central galaxy. The predicted velocity dispersion profile for the
central galaxy of MS2137--2353 is similar to the velocity dispersion
profile of NGC 6166, and in modeling the lensing cluster,
Miralda-Escud\'e (1995) also ruled out dark matter density profiles with
$\rho \propto r^{-1}$ in the inner region of the cluster.

To summarize the mass profile we derived reproduces our observation of
the kinematics in NGC 6166 and the surrounding cluster and is also
consistent with the X-ray mass determinations for Abell 2199 and with
observations of strong lensing in intermediate redshift clusters.

%%%%%%%%%%%%%%%%%%%%%%%%%%%%%%%%%%%%%%%%%%%%%%%%%%%%%%%%%%%%%%%%%%%%%%%%

\section{Discussion}
\label{discuss}

The cD galaxy NGC 6166 clearly shows a rise in $\sigma$ with projected
radius, nearly reaching the observed velocity dispersion of the galaxies
in the surrounding cluster. NGC 6166 is only the second cD galaxy with a
measured velocity dispersion profile that rises at large radii, after IC
1101 \citep{dressler}. A rising velocity dispersion profile is not seen
in the simulation by Dubinski (1998), in which he constructs a cD-like
galaxy through the continuous accretion of rotationally-supported
spiral-like galaxies.

The soft ($\alpha < 1$) uniform density core of NGC 6166's dark halo is
also inconsistent with cosmological $N$-body simulations with inert CDM
\citep{romeel}. Many authors have now begun to explore other forms of
cold dark matter, such as self-interacting, warm, or differently mutated
forms of dark matter \citep{spergel,yoshida2,romeel,colin,dalcanton}.
The most extreme cases of active dark matter produce flat ($\alpha=0$)
halo cores, but some fine-tuning of the dark matter interaction cross
section may be required to avoid creating a universe of perfectly
spherical halos \citep{jordi}. Perhaps the discrepancy will be resolved
with $\alpha$ being a function of scale \citep{jing}; though see
\citep{klypin}.

For the moment, the simulations are difficult to interpret. Not only is
the scatter in $\alpha$ poorly known, but the quoted values of $\alpha$
have been derived by fitting the halo profiles outside $r\simgt 0.03
R_{200}$\footnote{In our models $0.03 R_{200} \approx 50$ kpc!}
\citep{klypin}, and the consequences of the implied extrapolation of
the parameterized models to $r\rightarrow 0$ are ignored. Taylor \&
Navarro (2001) argue that the phase-space density profiles of dark
matter halos are well-approximated by power-laws over $0.01 R_{200}
\simlt r \simlt R_{200}$, and those authors extrapolate the power-law
phase-space density profile to $r=0$ to obtain $\alpha=0.75$ 
\citep[for NGC
6166 $\alpha=0.75$ implies $M/L_B=4.5$, too low by a factor of two,][]
{gerhard}. Including the effects of baryons will likely be
important on these scales as the cooling of baryonic matter will result
in adiabatic contraction \citep{blumenthal,rix,keeton}, which steepens
the inner profile of the halo \citep{alvarez}. It has also been
suggested that the transfer of angular momentum from the baryons to the
dark matter can soften the cores of halos \citep{weinkatz,klypin2}.
Furthermore, Weinberg (2001) suggests a number of dynamical processes
that can drive the cores of dark matter halos towards $\alpha\approx 0$.

Future simulations will hopefully resolve these problems either (1)
by using more physically motivated boundary conditions in the extrapolation
to $r\rightarrow 0$ \citep[cf.][]{taylor}, or (2) by resolving the physics
within $r \simlt 3\% R_{200}$. Because Weinberg (2001) and Nusser \&
Sheth (1999) suggest that the inner profiles of dark matter halos may be
sensitive to their initial conditions and histories of collapse and
merging, the simulations should ultimately (1) predict the cosmic
scatter in $\alpha$ (at a given halo mass), and (2) provide the means to
estimate the timescales of formation for dark matter halos of a given
mass and $\alpha$ \citep[instead of using the concentration
parameter;][]{bullock}.

%%%%%%%%%%%%%%%%%%%%%%%%%%%%%%%%%%%%%%%%%%%%%%%%%%%%%%%%%%%%%%%%%%%%%%%%

\section{Conclusions}
\label{summary}

We use the stellar kinematics of NGC 6166 and its halo to probe the mass
distribution in the core of the rich cluster Abell 2199. The projected
velocity dispersion along the major-axis of NGC 6166 rises steeply with
radius beyond the projected radius $R\approx 10$ kpc (where $\mu_B
\approx 23$ mag/arcsec$^2$). The velocity dispersion profile decreases
from about 300 \kms\ at $R=0$ kpc to 200 \kms\ at $R=2$ kpc (recall, the
seeing was $0\Sec 8$ FWHM $\equiv 0.45$ kpc), remains
flat to $R\approx 10$ kpc, and then steadily rises to $660\pm 120$ \kms
at $R \approx 60$ kpc (where $\mu_B \approx 26$ mag/arcsec$^2$), nearly
reaching the velocity dispersion of the cluster ($\sigma_{\rm
A2199}=775\pm 50$ \kms). There is also evidence for mild rotation in
intracluster stars beyond $R=20$ kpc ($|V/\sigma| = 0.3$), which is not
seen in the kinematics of the cluster members. Because we measure the
velocity dispersion gradient with high precision, we have constrained
the mass profiles of both the stellar and dark matter distributions,
under the assumptions of isotropy and spherical symmetry.

The primary results of our work indicate that the gravitational
potential of the cD halo is dominated by dark matter outside $R=20$ kpc.
Models in which the dark halo is represented by an $\alpha=1.5$
generalized-NFW density profile \citep[e.g.,][]{moore} are not able to
reproduce the observed dispersion profile. While the $\alpha=1$ NFW
profile is an improvement, the implied $M/L_B=1$ is not consistent with
what has been measured for other giant elliptical galaxies \citep[$8\simlt
M/L_B\simlt 12$;][]{gerhard}. In the models that utilize the
$\alpha=0$ generalized-NFW dark matter halo, one obtains $M/L_B=9$, a
more reasonable value.

The two-component mass models are consistent with the mass profiles
derived from X-ray observations \citep{markevitch,sidd} over the range
of radii for which our observations overlap. We also derived tangential
critical radii for our two-component mass models assuming Abell 2199
were placed at cosmological redshifts. The predicted tangential critical
radii are consistent with the radii of arcs in strong lensing clusters
at intermediate redshift.

Our results obtained for Abell 2199 have strong implications for the
study of dark matter halo profiles. The best-fit models are inconsistent
with the predictions of $N$-body simulations of cold dark matter.
However, we cannot yet verify the universality of dark matter halo
profiles. Simulations are needed to provide more robust estimates of
$\alpha$ at radii $r< 0.01 r_{200}$ \citep[see, e.g.,][]{klypin} and to
predict the scatter in $\alpha$. Observationally, we are obtaining
kinematical information for additional cD galaxies with the goal of
providing empirical constraints on the scatter in dark matter halo
properties.

With the possibility of observing cD galaxies to large radii using very
long ($\gg 10'$) slits, the task of background subtraction will become
easier and more accurate. Using a large database of cD galaxy
kinematics, with several long-slit position angles per galaxy, we can
explore more complicated models, in which both intrinsic flattening and
anisotropy are included. At the same time, the expansion of catalogs of
cluster galaxies \citep[e.g.,][]{rines}, and dramatic increases in the
spatial resolution of $X$-ray data, will enable us to better constrain
the inner and outer slopes of dark matter halo profiles, while allowing
us to avoid many of the simplifying assumptions. Such improvements may
soon lead to the confirmation or refutation of universal dark matter
density profiles, at least for the largest dynamically bound systems.

\acknowledgements

We would like to acknowledge helpful discussions with T. Lauer, D.
Eisenstein, A. Dressler, and G. Evrard, R. Dav\'e, F. van den Bosch, and
to thank the entire staff of the W.M. Keck Observatory for their
support. We also thank the anonymous referee for suggestions which
greatly improve the presentation of the material. AIZ acknowledges
support from NASA grants NAG 5-11108 and HF-01087.01-96A. SCT
acknowledges support from NASA grant HF-01125.01-99A and through grant
NAS 5-26555. JSM acknowledges support from NASA grant NAG 5-3529. This
research has made use of the NASA/IPAC Extragalactic Database (NED)
which is operated by the Jet Propulsion Laboratory, California Institute
of Technology, under contract with the National Aeronautics and Space
Administration.

%%%%%%%%%%%%%%%%%%%%%%%%%%%%%%%%%%%%%%%%%%%%%%%%%%%%%%%%%%%%%%%%%%%%%%%%

%%%%%%%%%%%%%%%%%%%%%%%%%%%%%%%%%%%%%%%%%%%%%%%%%%%%%%%%%%%%%%%%%%%%%%%%
%\clearpage
%\section{Figures}

\clearpage

\begin{deluxetable}{c c c c c c c c c c c}
\scriptsize
\tablewidth{0pt}
\tablecaption{Structural Parameters for the Stellar
Profile$^a$\label{tab:struct}}
\tablehead{
&&\multicolumn{2}{c}{$r^{1/4}$-law$^b$}&
\multicolumn{4}{c}{King Model$^c$}&
\multicolumn{3}{c}{Power-Law$^d$}\\
\colhead{Filter}&
\colhead{Correction$^e$}&
\colhead{$r_e$ (kpc)}&
\colhead{\rms\ (mag)}&
\colhead{$r_c$ (kpc)}&
\colhead{$r_{1/2}$ (kpc)}&
\colhead{$\Psi_0/\sigma^2$}&
\colhead{\rms\ (mag)}&
\colhead{$r_c$ (kpc)}&
\colhead{$\gamma$}&
\colhead{\rms\ (mag)}
}
\startdata
$B$& None & 61.3 & 0.073 & 3.84 & 102  & 10.25 & 0.071 & 2.35 & 1.22 & 0.034 \\
$R$& None & 47.5 & 0.058 & 3.58 & 46.2 &  9.50 & 0.069 & 2.45 & 1.26 & 0.048 \\
\\                                
$B$& Age  & 19.7 & 0.064 & 2.82 & 18.2 & 8.22 & 0.033 & 2.94 & 1.50 & 0.059 \\
$R$& Age  & 19.7 & 0.052 & 2.78 & 18.2 & 8.23 & 0.029 & 2.92 & 1.49 & 0.070 \\
\\                                
$B$& [Fe/H]& 34.2 & 0.064 & 3.30 & 31.6 & 8.85 & 0.042 & 2.66 & 1.35 & 0.046 \\
$R$& [Fe/H]& 34.2 & 0.053 & 3.28 & 31.6 & 8.86 & 0.051 & 2.63 & 1.35 & 0.055 \\
\enddata
\tablecomments{
$^a$To avoid the effects of the seeing disk, we used an inner fitting
radius of $r_{in}=2\Sec 5$. We also restricted the outer fitting radius
to $r_{out}=75$ kpc because that radius is also the outer radius for the
stellar kinematics.
$^b$The $r^{1/4}$-law has been fit so that the resultant half-light
(effective) radii can be compared to the half-light radii derived using
the other profiles.
$^c$The King (1966) profile has two free parameters: $r_c$, the core
radius, and $\Psi_0/\sigma^2$, the truncation energy \citep[see][]{bt87}.
$^d$The power-law, of the form $\rho\propto [1+(r/rc)^2]^{-\gamma}$, has
two free parameters as well. Note that because the best-fit power-law
profile has $\gamma<2$, the total mass/luminosity is divergent and there
can be no half-light radius.
$^e$The color gradient may be interpreted as a spatial variation in the
stellar population ages and/or metallicities (see text), and we use this
interpretation to explore corrections to the radial variation in
$(M/L_*)$. Even though the color gradient in NGC 6166 is small (see
text), the inferred half-mass radii are very sensitive to the
correction.
}
\end{deluxetable}

\begin{deluxetable}{r r r r r r r}
\scriptsize
\tablewidth{0pt}
\tablecaption{Major-Axis Kinematics for NGC 6166
\label{tab:kinprof}}
\tablehead{
\colhead{R}&
\colhead{$\Delta R$}&
\colhead{$\sigma$}&
\colhead{$\delta_\sigma$}&
\colhead{$V$}&
\colhead{$\delta_V$}&
\colhead{$\chi^2$}\\
\colhead{(kpc)}&
\colhead{(kpc)}&
\colhead{(\kms)}&
\colhead{(\kms)}&
\colhead{(\kms)}&
\colhead{(\kms)}&
\colhead{}\\
\colhead{(1)}&
\colhead{(2)}&
\colhead{(3)}&
\colhead{(4)}&
\colhead{(5)}&
\colhead{(6)}&
\colhead{(7)}
}
\startdata
-17.16 &   6.29  &   404.7 &   47.5 &   -40.4 &   34.1 &  1.36\\
-12.47 &   2.61  &   326.1 &   59.1 &    14.0 &   27.1 &  1.21\\
 -9.74 &   2.37  &   267.1 &   26.6 &   -12.3 &   15.8 &  1.14\\
 -7.66 &   1.31  &   243.6 &   23.7 &   -12.7 &   13.7 &  1.07\\
 -6.35 &   0.83  &   207.1 &   26.6 &   -19.8 &   15.1 &  1.17\\
 -4.69 &   2.02  &   219.9 &   13.5 &   -10.1 &    7.6 &  1.49\\
 -2.97 &   0.95  &   216.7 &   12.4 &   -10.9 &    7.0 &  1.30\\
 -1.96 &   0.59  &   200.3 &   14.5 &     0.9 &    7.5 &  1.45\\
 -1.25 &   0.36  &   215.4 &   14.4 &     4.2 &    7.9 &  1.27\\
 -0.71 &   0.24  &   264.9 &   14.3 &     0.7 &    8.9 &  1.18\\
 -0.24 &   0.24  &   305.7 &   18.8 &    -2.6 &    5.5 &  1.35\\
  0.24 &   0.24  &   289.0 &   16.9 &     0.9 &    6.7 &  1.27\\
  0.71 &   0.24  &   247.2 &   15.3 &    -3.9 &    9.7 &  1.44\\
  1.25 &   0.36  &   209.5 &   13.1 &     4.5 &    8.3 &  1.39\\
  1.90 &   0.47  &   208.2 &   13.7 &    -9.2 &    8.2 &  1.29\\
  2.73 &   0.71  &   217.9 &   14.2 &     4.1 &    7.8 &  1.18\\
  3.80 &   0.24  &   226.4 &   15.0 &   -18.7 &    7.9 &  1.04\\
  3.86 &   1.07  &   216.2 &   15.4 &     6.7 &    8.4 &  1.20\\
  4.27 &   0.47  &   240.8 &   10.1 &    -7.1 &    5.3 &  1.77\\
  4.93 &   0.59  &   220.1 &    9.3 &    -2.8 &    4.6 &  1.46\\
  5.40 &   1.54  &   247.0 &   16.6 &     1.5 &    7.8 &  1.16\\
  5.58 &   0.47  &   197.3 &   12.0 &     4.0 &    5.0 &  1.34\\
  6.17 &   0.47  &   191.6 &   11.8 &    17.1 &    4.8 &  1.22\\
  7.01 &   0.95  &   209.5 &   11.3 &     4.0 &    4.2 &  1.48\\
  8.01 &   3.21  &   233.9 &   19.6 &     6.1 &    7.6 &  1.12\\
  8.19 &   1.19  &   213.0 &   11.8 &    -5.6 &    4.2 &  1.49\\
  9.44 &   1.07  &   221.7 &   14.8 &   -10.4 &    5.5 &  1.36\\
 10.69 &   1.19  &   282.7 &   15.6 &     8.1 &    6.6 &  1.41\\
 12.05 &   1.31  &   262.9 &   16.1 &   -27.7 &    8.0 &  1.18\\
 13.42 &   1.42  &   248.0 &   16.8 &     5.3 &   10.6 &  1.36\\
 15.67 &   2.85  &   275.0 &   15.3 &    -4.7 &    9.4 &  1.37\\
 18.64 &   2.85  &   316.1 &   17.4 &    -5.3 &   11.7 &  1.25\\
 27.49 &   2.97  &   351.1 &   29.3 &    38.1 &   19.5 &  1.39\\
 30.81 &   3.44  &   445.6 &   60.5 &    45.0 &   37.9 &  1.35\\
 34.91 &   4.75  &   490.9 &   58.6 &    53.1 &   29.9 &  1.72\\
 39.30 &   4.04  &   529.1 &   73.9 &   114.3 &   47.3 &  1.32\\
 57.77 &  24.34  &   663.0 &  121.2 &   232.6 &   61.8 &  2.38\\
\enddata
\end{deluxetable}

\end{document}